\documentclass{mn2e}
\usepackage{psfig}

\newcommand{\Msol}{\mbox{$\rm M_{\odot}$}}
\newcommand{\hMsol}{\mbox{$h^{-1}\Msol$}}
\newcommand{\keV}{\mbox{\rm keV}}
\newcommand{\kpc}{\mbox{\rm kpc}}
\newcommand{\hkpc}{\mbox{$h^{-1}\kpc$}}
\newcommand{\Mpc}{\mbox{\rm Mpc}}
\newcommand{\hMpc}{\mbox{$h^{-1}\Mpc$}}

\title[Cooling and preheating in X-ray clusters]
{The effect of cooling and preheating on the X-ray properties of 
clusters of galaxies}
\author[O. Muanwong et~al.]
{Orrarujee ~Muanwong,$^1$\thanks{Email: O.Muanwong@sussex.ac.uk} 
Peter A. ~Thomas$^1$, Scott T. ~Kay$^1$
and Frazer R. ~Pearce$^2$ \\
$^1$ Astronomy Centre, CPES, University of Sussex, Falmer, Brighton BN1 9QJ\\
$^2$ Physics and Astronomy Department, University of Nottingham,
Nottingham, NG7 2RD}

\pagerange{\pageref{firstpage}--\pageref{lastpage}}
\pubyear{2002}
\date{\today}
\begin{document}
\journal{Preprint astro-ph/0205137} 
 
\maketitle

\label{firstpage}

\begin{abstract}
We calculate X-ray properties of present-day galaxy clusters from
hydrodynamical cosmological simulations of the $\Lambda$CDM cosmology
and compare these with recent X-ray observations.  Results from three
simulations are presented, each of which uses the same initial
conditions: a standard adiabatic, {\it Non-radiative}\ model, a {\it
Radiative} model that includes radiative cooling of the gas, and a
{\it Preheating} model that also includes cooling but in addition
impulsively heats the gas prior to cluster formation.  At the end of
the simulations, the global cooled baryon fractions in the latter two
runs are 15 per cent and 0.4 per cent respectively which bracket the
recent result from the {\it K}-band luminosity function.  We construct
cluster catalogues which consist of over 500 clusters and are complete
in mass down to $1.18\times10^{13} h^{-1}\Msol$.  While clusters in
the {\it Non-radiative} model behave in accord with the self-similar
picture, those of the other two models reproduce key aspects of the
observed X-ray properties: \emph{the core entropy, temperature-mass
and luminosity-temperature relations are all in good agreement with
recent observations}.  This agreement stems primarily from an increase
in entropy with respect to the {\it Non-radiative} clusters.  Although
the physics affecting the intra-cluster medium is very different in
the two models, the resulting cluster entropy profiles are very
similar.

\end{abstract}


\section{Introduction}
Because of numerical limitations, most previous simulations of
clusters of galaxies have been adiabatic, resulting in approximate
self-similar scaling for cluster properties.  However, these models
are unphysical in that the cooling time of the gas in the central
regions of the clusters is less than their age; also observed
clusters do not scale self-similarly.  In this paper, we generate mock
cluster catalogues from simulations that include radiative cooling and
we show that these can reproduce the observed scaling relations.

The observed X-ray luminosity-temperature (${L_X - T_X}$) relation has
a slope as large as 3 (e.g.~Edge \& Stewart 1991), steeper than the
predicted value of 2 from self-similarity.  Cooling flows, which are
known to exist in a large population of clusters (Allen \& Fabian
1998), are one cause of this as the more massive, cooling flow
clusters have an excess of luminosity.  However, when the cooling flow
components are corrected for, the observed slopes are still
significantly larger than 2 (Allen \& Fabian 1998; Markevitch 1998).
The departure from self-similarity is even more prominent in low mass
clusters or groups (Ponman et al. 1996; Xue \& Wu 2000) where the
slope can be as steep as 8.  These departures from self-similarity can
help us to learn about the history of the intra-cluster medium (ICM).

The ${L_X - T_X}$ relation predicted from self-similarity, when
bremsstrahlung is assumed to dominated the emission, can be written as
$L_X \propto f^2_{\rm gas}(1+z_{\rm form})^{3/2}T^2$ where $f_{\rm
gas}$ is the gas mass fraction and $z_{\rm form}$ is the formation
redshift. Mohr, Mathiesen \& Evrard (1999) and Arnaud \& Evrard (1999)
present evidence that the intracluster medium is more extended in
low-mass clusters so that $f_{\rm gas}$ is a mildly-increasing
function of temperature, but the latter also stress that this is
statistically inconclusive due to the fact that their sample is not
homogeneous and the results are model-dependent.  

One explanation for the above observations is that galaxy formation is
more efficient in poor groups than clusters.  David et al. (1990) show
that the ratio between the gas and stellar mass increases with gas
temperature. They argue that as mergers take place to form massive
clusters, the gas is heated and this suppresses galaxy formation.
Bryan (2000) compiled observed mass fractions in groups and clusters
from various studies.  By assuming variations of galaxy formation
efficiencies and applying it to a simple model, he successfully
predicted the $L_X - T_X$ relation.  This motivates our {\it
Radiative} simulation in which gas in removed from the ICM
by radiative cooling.  We are somewhat fortunate in that the finite
resolution of the simulation limits the amount of cooling (or else the
vast majority of the gas would have cooled to low temperatures) and
leads to a similar amount of star formation as Bryan suggests is
required to reproduce the observations.  

Kaiser (1991) and Evrard \& Henry (1991) suggest an alternative way of
expelling gas from clusters via energy injection at early times,
perhaps feedback from galaxy formation.  The gas does not have to be
removed from the cluster entirely but only from the central regions
that dominate the X-ray emission.  Observational evidence is provided
by Ponman, Cannon \& Navarro (1999) who show that the surface
brightness profiles of cool and hot systems are not scaled version of
each other and that the core entropies of low temperature clusters are
higher than can be achieved by gravitational collapse alone.  It
appears that the core entropy approaches a certain value,
approximately 100\,$h^{-1/3}$\,keV\,cm$^2$, at low temperature that
they designate the `entropy floor'.

For this preheating model to work, it has to happen in an optimal way,
i.e.~the right amount of energy has to be injected at the right time.
There have been various studies which suggest that if an energy is
injected well before cluster collapse, 0.3 keV per particle is
required (Lloyd-Davies, Ponman \& Cannon 2000).  If energy injection
takes place internally, i.e.~within collapsed halos, more energy is
required, 1-3 keV per particle (Lowenstein 2000; Wu, Fabian \& Nulsen
2000; Bower et al. 2001).  The efficiency of energy injection depends
on the density of the IGM: the higher the density, the more energy is
required to achieve the same entropy level.  These results motivate
our {\it Preheating} simulation in which, in addition to radiative cooling,
the gas is preheated by raising the specific energy by 1.5\,keV at a
redshift of 4.

It has been shown that there is a need of some form of feedback energy
into the ICM in order to get a suitable cooled baryon fraction (White
\& Frenk 1991, Cole 1991, Blanchard, Valls-Gabaud \& Mamon
1992). Without such process, too much gas would be cooled and
converted into stars at high redshift.  As halos at high redshift are
small and dense, cooling becomes very efficient. A vast fraction of
gas is cooled into small systems leaving only a small amount of hot
gas in the halos to form galaxies at later times, the so-called
`cooling catastrophe' (White 1992).  The picture of this process is
clearly in contradiction with observations as large amounts of hot
baryons are observed in X-ray clusters at the present day.  

Unfortunately, an accurate determination of the cooled gas fraction in
clusters of galaxies is hard to obtain.  From recent observations of
the $K$-band luminosity function, Balogh et~al.~(2001) show that the
global cooled baryon fraction is only about 5 per cent. Using a sample
of clusters compiled by Roussel et~al.~(2000) and Carlberg
et~al.~(1996), they estimated a cooled fraction of around
20 per cent for systems with $kT=$1\,keV, decreasing to around 10 per
cent at $kT=$10\,keV.  Note, however, that these results rely on
extrapolating the observations out to the virial radii (by as much as
a factor of 5 in radius for low-temperature systems).  These
observations do not preclude excess cool material in a form other than
stars, such as neutral or molecular gas.  Indeed, the models of Bryan (2000)
and Wu \& Xue (2002b) both required a cooled gas fraction in clusters
of about twice the above in order to reproduce the X-ray scaling relations.

Despite this uncertainty in the cooled gas fraction in real clusters,
our {\it Radiative} and {\it Preheating} simulations are likely to
bracket the true value.  We will show below that in all other
resepects the clusters in our simulations reproduce the X-ray
properties of clusters extremely well.  This agreement stems from the
entropy profiles which are similar in the two simulations and much
shallower than in a {\it Non-radiative} simulation.

In Section 2, we describe the simulations and explain how we construct
our simulated cluster catalogues.  The properties of the ICM are
presented in Section 3, and the temperature-mass and
luminosty-temperature scaling relations are discussed in Section
4. Finally we summarise our conclusions in Section 5.

\section{Methodology}
\subsection{Simulation details}

Simulation data were generated using a parallel implementation of the
{\sc hydra} code (Couchman, Thomas \& Pearce 1995; Pearce \& Couchman
1997), which uses the adaptive particle-particle/particle-mesh
(AP$^3$M) algorithm to calculate gravitational forces (Couchman 1991)
and Smoothed Particle Hydrodynamics (SPH) to model hydrodynamical
forces.  Our implementation of SPH is similar to that used by and
discussed in Thacker \& Couchman (2001).

Results are presented assuming a flat, low-density cosmology, setting
the density parameter $\Omega_0=0.35$, Hubble constant
$h=0.71$\footnote{We define $h=H_0/100$\,km\,s$^{-1}$Mpc$^{-1}$},
cosmological constant $\Omega_{\Lambda}=\Lambda/3H_0^{2}=0.65$, baryon
density $\Omega_{\rm b}=0.019\,h^{-2}$, cold dark matter power spectrum
shape parameter $\Gamma = 0.21$ and normalization $\sigma_8 =
0.9$. The initial density field was realised by perturbing
4,096,000 ($160^3$) particles each of dark matter and gas from a regular cubic
mesh of comoving length, 100$\hMpc$. Thus, dark matter and gas
particle masses are approximately $2.1 \times 10^{10} \hMsol$ and $2.6
\times 10^{9} \hMsol$ respectively. Each simulation was then evolved
to $z=0$, typically taking around 2000 steps, using 64 processors on
the Cray T3E at the Edinburgh Parallel Computing Centre.  The
gravitational softening length was fixed at $\epsilon=50 \hkpc$ in
comoving co-ordinates (equivalent Plummer value) until $z=1$, after
which it was fixed in physical co-ordinates at $\epsilon=25 \hkpc$
until $z=0$. This choice prohibited unwanted relaxation effects (see
below).

In this paper we study 3 simulations that differ only in the manner in
which the gas particles are cooled and heated:
\begin{description}
\item{\it Non-radiative}: This simulation does not contain any
radiative cooling, and heating occurs solely by adiabatic compression
and shock-heating.  As discussed previously (Thomas et al. 2001;
Muanwong et al. 2001, hereafter T2001 and M2001, respectively) this
run does not provide an accurate physical description of clusters
since gas with short cooling times cannot cool, but serves as a useful
model for comparison and to test numerical effects.
\item{\it Radiative}: In this run the gas was allowed to cool
radiatively, using cooling tables from Sutherland \& Dopita (1993). We
adopted a global gas metallicity of $Z=0.3(t/t_0)Z_{\odot}$, where
$t/t_0$ is cosmic time in units of the present value. Material which
had cooled (identified as all gas particles with overdensities
$\delta > 1000$ and temperatures $T < 1.2 \times 10^{4}$K) was
identified on each step, and groups of 13 particles within a softening
length were merged into single collisionless particles (hereafter
referred to as `galaxy fragments'). Fragments could subsequently
accrete more cooled gas (within a softening length) but not merge with
other fragments.  We do not discuss the properties of these galaxy
fragments in this paper.
\item{\it Preheating}: This is identical to the {\it Radiative} run,
except that all gas particles were impulsively heated by adding 1.5\,keV
of thermal energy at $z=4$, corresponding to an increase in
temperature of approximately $1.2\times 10^{7}$K.
\end{description}
As we shall see, both the {\it Radiative} and {\it Preheating}
runs reproduce the
observed cluster scaling relations but contain
vastly different amounts of cooled gas.

\subsection{Cluster identification}
\label{subsec:clusid}

In this paper, we present results only for clusters at $z=0$.
\footnote{For clarity, we choose not to discriminate between groups 
and clusters and use only the latter term.}  Clusters are identified
using an identical procedure to that used by M2001, as described below.

First, we create a minimal-spanning tree of all dark matter particles
whose density exceeds 317 times the mean dark matter density in the
box.  (For a spherical top-hat model, this is equal to the mean density
within virialised regions, although the precise density used does not
matter at this stage.)
We then define a maximum linking length equal to
$0.5\,(317)^{-1/3}$ times the mean inter-particle separation and use
this to split the minimal-spanning tree into clumps of particles that
serve as potential sites of clusters.  The centre of each clump is
defined as the position of the densest dark matter particle, then a
sphere is grown around all particles until radii are found that enclose 
average overdensities of 111 (i.e. 317\,$\Omega_0$), 200, 500, 1000 and 2500 
relative to the critical density, $\rho_{\rm cr}=3H_0^2/8\pi G$. 
Clumps are written into the
cluster catalogues provided that they contain a mass equivalent to at
least 500 particles each of gas and dark matter within these radii,
and that their centres are not located within a more massive cluster.

As we are interested in the X-ray properties of clusters, we did
consider using gas particles rather than dark matter particles to
define the cluster centre.  However, we decided against this for two
reasons: firstly, in runs with radiative cooling the densest gas
particle was often located in the outskirts of the cluster as defined
by the dark matter; secondly, it made it hard to compare clusters from
different simulations that had similar dark matter distributions
but very different gas ones.

The catalogues are complete in mass down to $1.18\times10^{13}
\hMsol$.  Our final catalogues for each simulation have of the
order of 530, 460, 350, 250 and 150 clusters within the overdensities
mentioned above, respectively.  Although we do not report on them in
this paper, we have also created cluster catalogues for a large number
of higher redshifts, and all the catalogues are available
on-line at http://virgo.sussex.ac.uk/.

\subsection{Numerical heating}
\label{sect:numerics}

\begin{figure}
\psfig{file=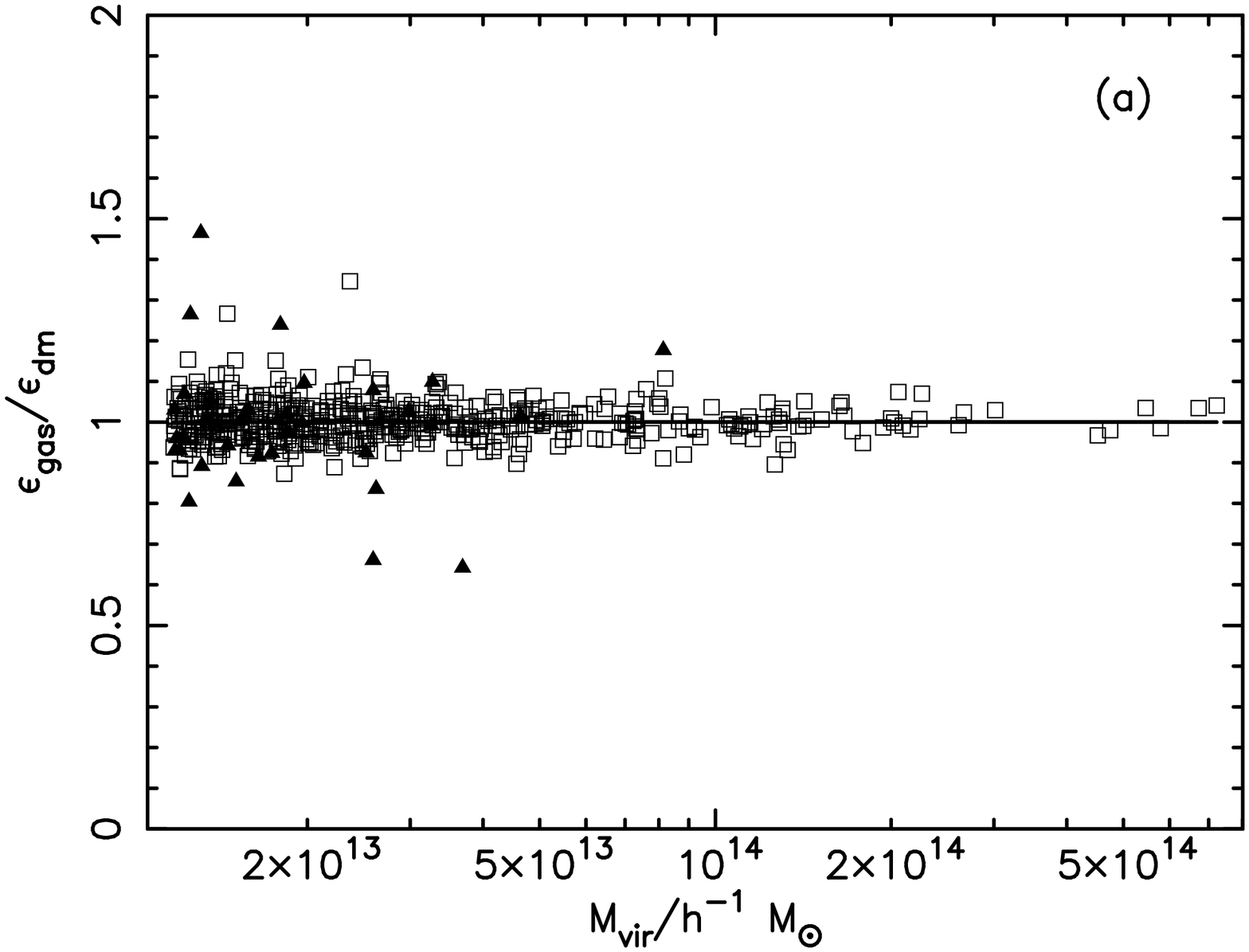,angle=270,width=8.5cm}
\psfig{file=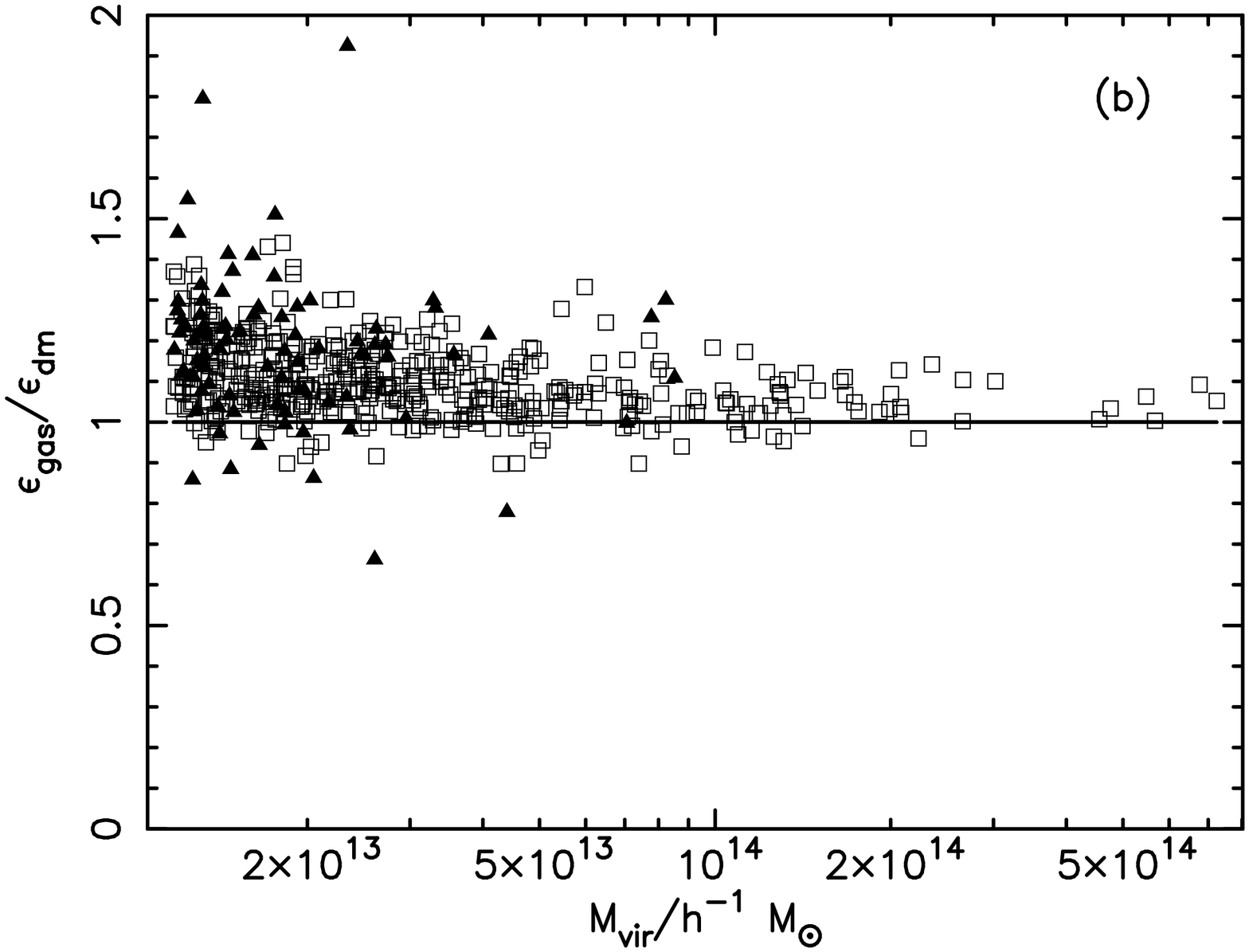,angle=270,width=8.5cm}
\caption{Ratios between the specific energy of the gas and
the dark matter within the virial radii of clusters in the (a) {\it
Non-radiative} and (b) {\it Radiative} simulations.  The filled
traingles correspond to clusters that show significant velocity
substructure, as described in the text.}
\label{fig:tratio}
\end{figure}

When undertaking an $N$-body simulation, it is crucial to use an
appropriate gravitational softening length (see, for example, T2001).
Ours is chosen following the Thomas \& Couchman (1992) criterion such
that the smallest clusters in all simulations presented here have
2-body relaxation times of at least five times the age of the
Universe. Also, for a metallicity $Z=0$, the mass of each dark matter
particle is approximately equal to the critical mass, as estimated by
Steinmetz \& White (1997), above which there would be enough spurious
numerical heating to suppress cooling within the haloes.  This is the
most conservative case, since $Z>0$ when the first resolved
haloes form in the simulations.

As a check on the degree of numerical heating in our simulations, we
plot in Fig.~\ref{fig:tratio} the ratio of the total specific energy
of the gas, $\epsilon_{\rm gas}$, to that of the dark matter,
$\epsilon_{\rm dm}$, as a function of mass, $M_{\rm vir}$, contained
within the virial radius, $r_{\rm vir}$, (defined as the radius which
encloses a mean overdensity of 111 times the critical density, in
accord with Section~\ref{subsec:clusid}). The specific energy of the
gas consists of the kinetic and thermal contributions of hot gas
particles (throughout this paper we use the term `hot gas' to mean
gas particles with temperatures in excess of $10^5$K) with masses
$m_i$, speeds $v_i$ (relative to the cluster mean) and temperatures
$T_i$:
\begin{equation}
\epsilon_{\rm gas} = 
  {\sum_im_i\big({1\over2}v_i^2+{3kT_i\over2\mu m_{\rm H}}\big)
  \over\sum_im_i},
\label{equ:epsgas}
\end{equation}
where $\mu m_{\rm{H}}$ is the mean molecular mass which we take to be
$10^{-24} \rm{g}$ for a cosmic mix of elements. (Note we assume
the ratio of specific heats, $\gamma=5/3$, corresponding to 
a monatomic ideal gas.) The specific energy of
the dark matter particles can similarly be written as
\begin{equation}
\epsilon_{\rm{dm}} = {\sum_i{1\over2}m_iv_i^2\over\sum_im_i}.
\label{equ:epsdm}
\end{equation}

In the {\it Non-radiative} simulation, the ratio of specific energies
(top panel of Fig.~\ref{fig:tratio}) is independent of cluster mass.
The outliers that have the highest and lowest values of the energy
ratio in the figures are subclumps which are falling into large
neighbouring clusters.  We illustrate this by using solid triangles to
denote clusters where the mean relative velocity of the gas and dark
matter is more than 0.1 times the velocity dispersion of the cluster
(and in fact one of the open squares that lie above the mean relation
designates a cluster that only just failed this cut).  When radiative
cooling is included (lower panel), the gas component is found to have
a higher specific energy than the dark matter and the ratio increases
with decreasing mass; similar results are found for the {\it
Preheating} simulation.  This result shows that radiative cooling
leads to mass-deposition and, paradoxically, a heating of the residual
ICM through adiabatic compression, a result first demonstrated in a
cosmological simulation by Pearce et al.~(2000; although see also
Knight \& Ponman 1997). In any event, the difference between the two
panels shows that physical heating processes overwhelm numerical ones
in our simulations.

\begin{figure}
\psfig{file=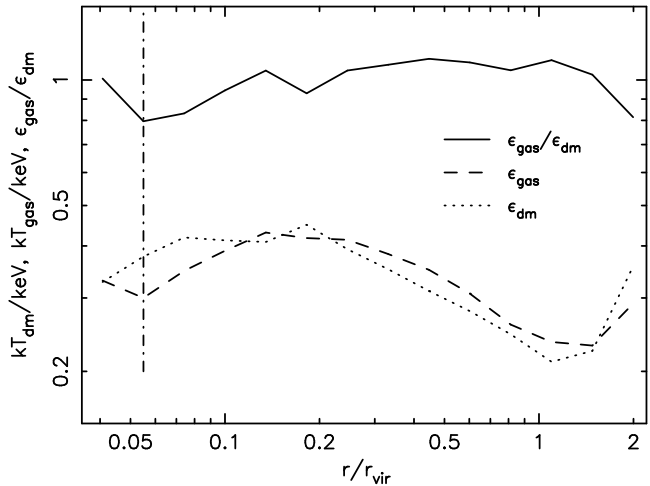,angle=270,width=8.5cm}
\caption{Averaged profiles of specific energy of the gas,
$\epsilon_{\rm gas}$, and the dark matter, $\epsilon_{\rm dm}$, and
the ratio between the two, for the 10 least massive clusters in the
{\it Non-radiative} simulation.  The units are defined in the usual
manner of X-ray observations in $kT$ units, such that $kT={2\over3}\mu
m_{\rm H}\epsilon$.The vertical line indicates the softening length.}
\label{fig:enproflow}
\end{figure}

To further test our results, we show in Fig.~\ref{fig:enproflow}
average energy profiles for the 10 smallest clusters in the {\it
Non-radiative} simulation.  The specific energies of the gas and dark
matter follow each other well at all radii and show no evidence for an
increase in gas specific energy over that of the dark matter near the
centres of the clusters.

\section{Properties of the intracluster medium}
\subsection{Baryon fractions}
\label{sect:barydist}

\begin{figure}
\psfig{file=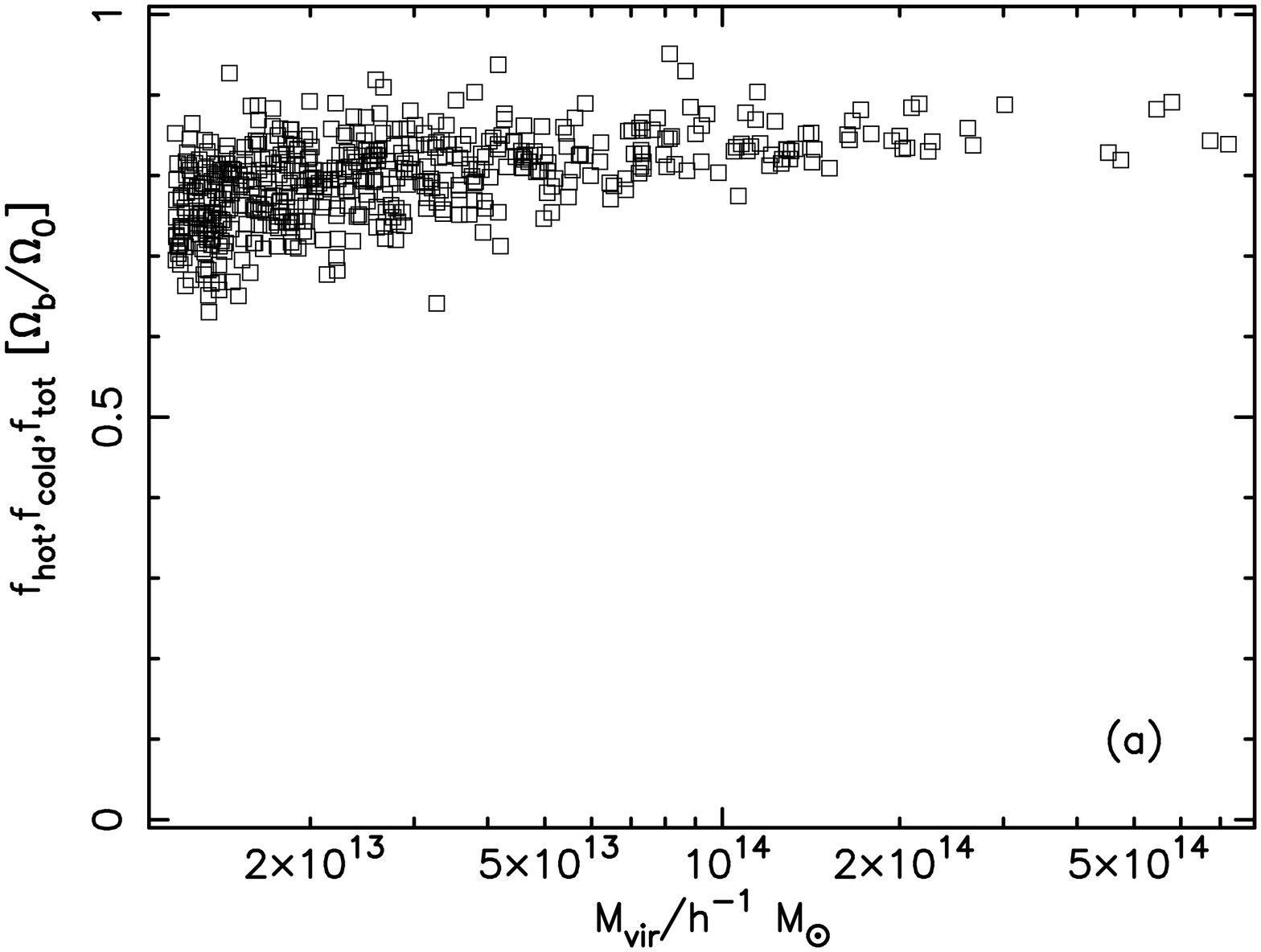,angle=270,width=8.5cm}
\psfig{file=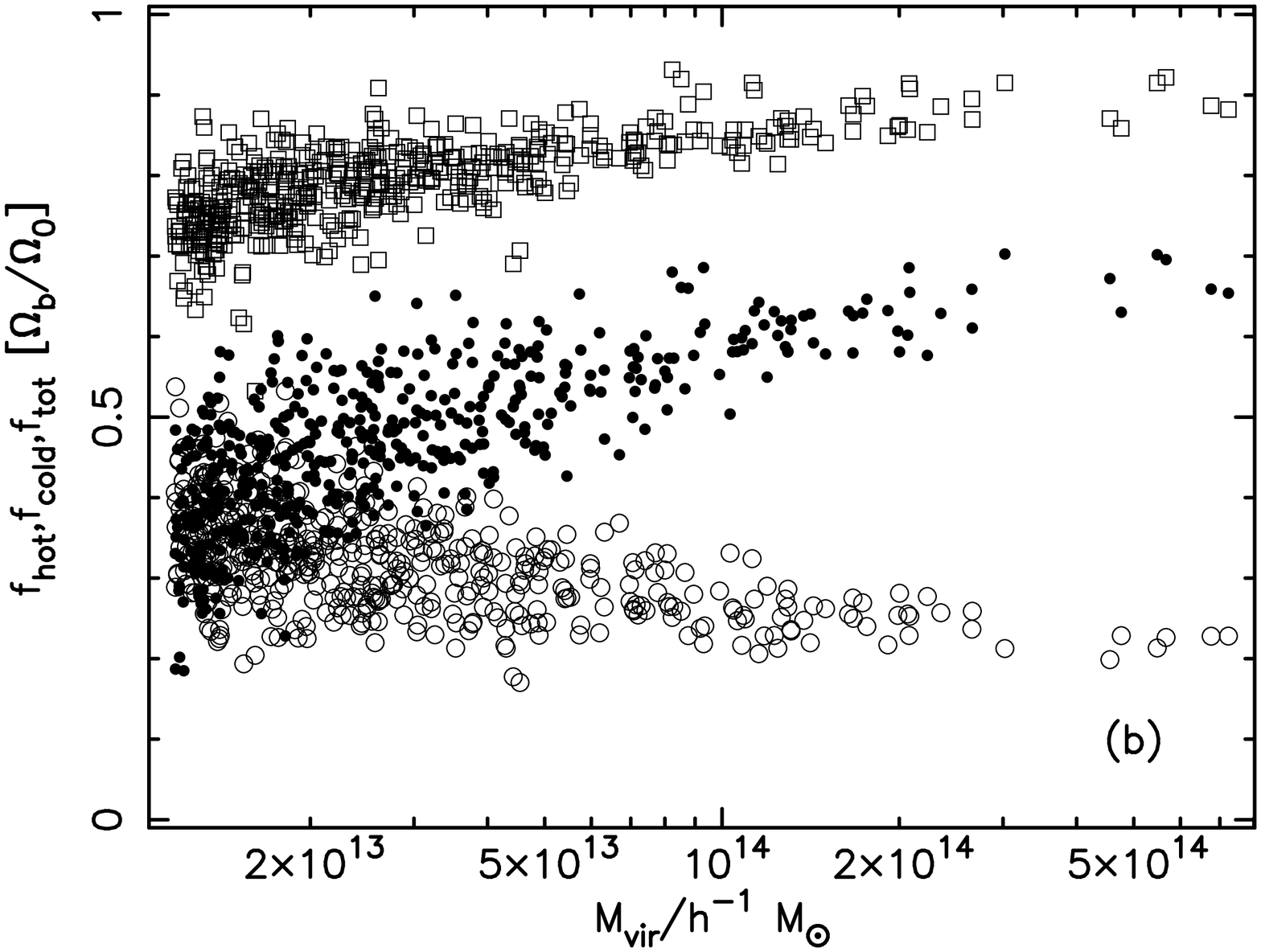,angle=270,width=8.5cm}
\psfig{file=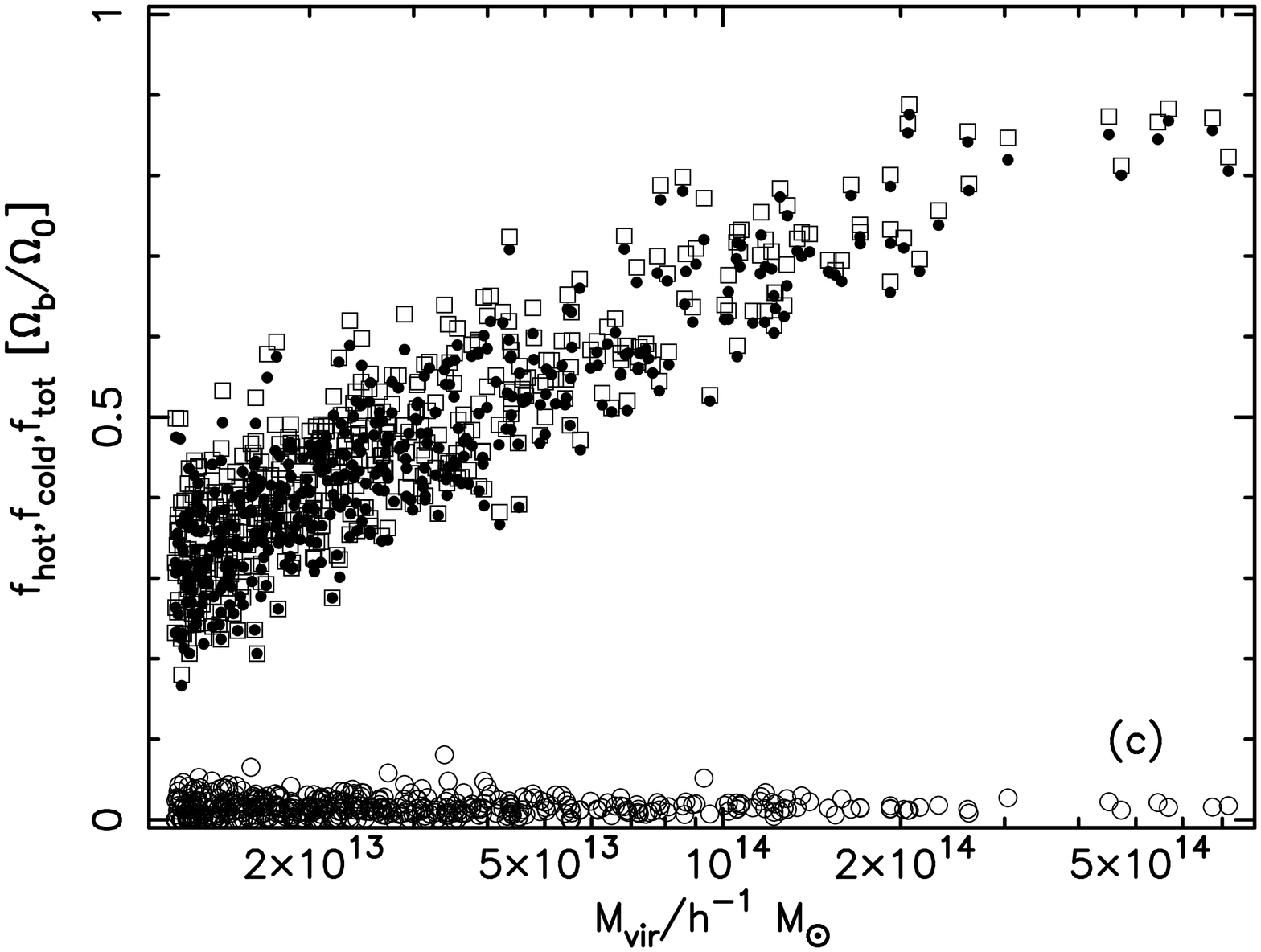,angle=270,width=8.5cm}
\caption{Mass fractions of hot (solid circles), cold (open circles)
gas and the total fractions (squares) within the virial radius of
clusters in the (a) {\it Non-radiative}, (b) {\it Radiative} and 
(c) {\it Preheating} simulations.}
\label{fig:fbary1}
\end{figure}

In this Section, we look at the relative distribution of baryons and
dark matter in the simulated clusters.  Panel (a) in Fig.~\ref{fig:fbary1}
shows the baryon fraction within the virial radius, in units of the
global baryon fraction, for the {\it Non-radiative} simulation.
Note that, even for the largest clusters in our catalogues, the ICM is
more extended than the dark matter. This is because, for a given
specific energy (see Fig.~\ref{fig:tratio}), the gas is better
able to support itself in the gravitational potential than the dark matter. 
The effect is more pronounced in smaller clusters although this may be
an artefact of poorer resolution in these objects.

In the {\it Radiative} run (Fig.~\ref{fig:fbary1}, panel b), the total
baryon fraction in the clusters is almost unchanged.
However, the mass fraction of hot gas (i.e.\ the hot
intracluster medium that gives rise to X-ray emission) is much
reduced, especially in the low-mass clusters.
The {\it Preheating} run (Fig.~\ref{fig:fbary1}, panel c) also
reduces the hot gas fraction in the centres of clusters, but it does
so by a different mechanism.  Instead of turning hot gas into cooled
gas (which term we take to include stars and galaxies), it heats it
up and expels it from the cluster core.  In low-mass clusters, the gas
is expelled from the cluster altogether resulting in a reduced baryon
fraction within the virial radius, whereas in high-mass clusters it is
merely redistributed to larger radii but remains within the cluster.

Neither the {\it Radiative} nor the {\it Preheating} runs give the
correct fraction of cooled gas.  The global cooled fraction in the box
is 15 per cent for the former and 0.4 per cent for the latter.
However, we will show later that the X-ray properties of the clusters
from the two simulations are very similar because they both have
similar entropy profiles.  Only when we look at the outer parts of
clusters can the two be easily distinguished.

\subsection{The entropy of intracluster gas}
\label{sec:entropy}

For a given gravitational potential, the spatial distribution of the
ICM in hydrostatic equilibrium is determined by its entropy 
(plus one normalisation condition such as the pressure at, or the total 
mass within, the virial radius).
Higher entropies correspond to more extended gas distributions and, in
particular, a high core entropy leads to a reduction in the gas
density near the centre of the cluster and hence a much reduced X-ray
luminosity.

In this paper, we follow the usual practice in observational X-ray
papers and work not with entropy but with the closely-related quantity
\begin{equation}
s_X(r) = \frac{kT_X(r)}{n_e(r)^{2/3}},
\label{eq:entropy}
\end{equation}
where $n_e$ is the electron density and $T_X$ is the 
emission-weighted X-ray temperature,
\begin{equation}
T_X = \frac{\Sigma_i m_i\rho_i\Lambda_{\rm soft}(T_i,Z)T_i}
          {\Sigma_i m_i\rho_i\Lambda_{\rm soft}(T_i,Z)}.
\label{eq:ktx}
\end{equation}
Here $m_i$, $\rho_i$ and $T_i$ are the mass, density and temperature
of the hot gas particles that contribute to the X-ray emission, $Z$ is
the metallicity, and $\Lambda_{\rm soft}$ is the cooling function from
Raymond \& Smith (1977) in the soft band, 0.3--1.5\,keV. (For a
fully-ionized plasma with a cosmic distribution of Helium, then
$n_e\approx0.88\rho/m_{\rm H}$.)

In the absence of radiative cooling, the entropy of the gas can only
increase through shock-heating associated with mergers and accretion.
Within any given cluster, there will be a range of entropies, with a
positive entropy gradient from the centre outwards (this comes about
both because shock-heating is more effective for material accreted
later and because any other distribution is convectively unstable).
For a self-similar cluster population for which we measure $s_X$ at
some fixed fraction of the virial radius (and hence the same value of
$n_e$ for each), then $s_X\propto T_X$.  In a real cluster population
this proportionality will not be exact because the profiles are not
exactly self-similar (the `concentration' varies with mass---see
e.g. Navarro, Frenk \& White 1997; T2001), but it will 
nevertheless hold to good approximation. 


However, this self-similarity does not seem to extend down to low-mass
clusters and groups.  Ponman et al.~(1999) and Lloyd-Davies et
al.~(2000) found that when measuring the entropy at a fiducial radius
of $0.1\,r_{\rm vir}$, which they termed the `core entropy', hot clusters
appeared to follow the prediction from self-similarity whereas cool
clusters lay above it. (We note that the term core entropy does not
imply that clusters show a constant entropy in the innermost
regions---they do not.  It is used in a loose term to define the
entropy outside the central cooling region but still near the centre
of the cluster.)  Although the observations are not very precise, a
rough interpretation is that the core entropy flattens off and
approaches a constant value, termed the `entropy floor', in small
systems.  Ponman et al. (1996) attributed this effect 
to preheating of the inter-galactic medium before cluster formation
(and thus motivated the {\it Preheating} model in this paper).

In this Section, we calculate the core entropy of our clusters in a
similar way as in the above observations.  We average the value of
$s_X$ in a spherical annulus centered on $0.1\,r_{\rm vir}$ and of
width $\Delta r=0.02\,r_{\rm vir}$ (there are various other ways of
doing this averaging but they all give similar results).  We note that
$0.1\,r_{\rm vir}$ is about twice the softening length for the
smallest clusters in our catalogues.

\begin{figure}
\psfig{file=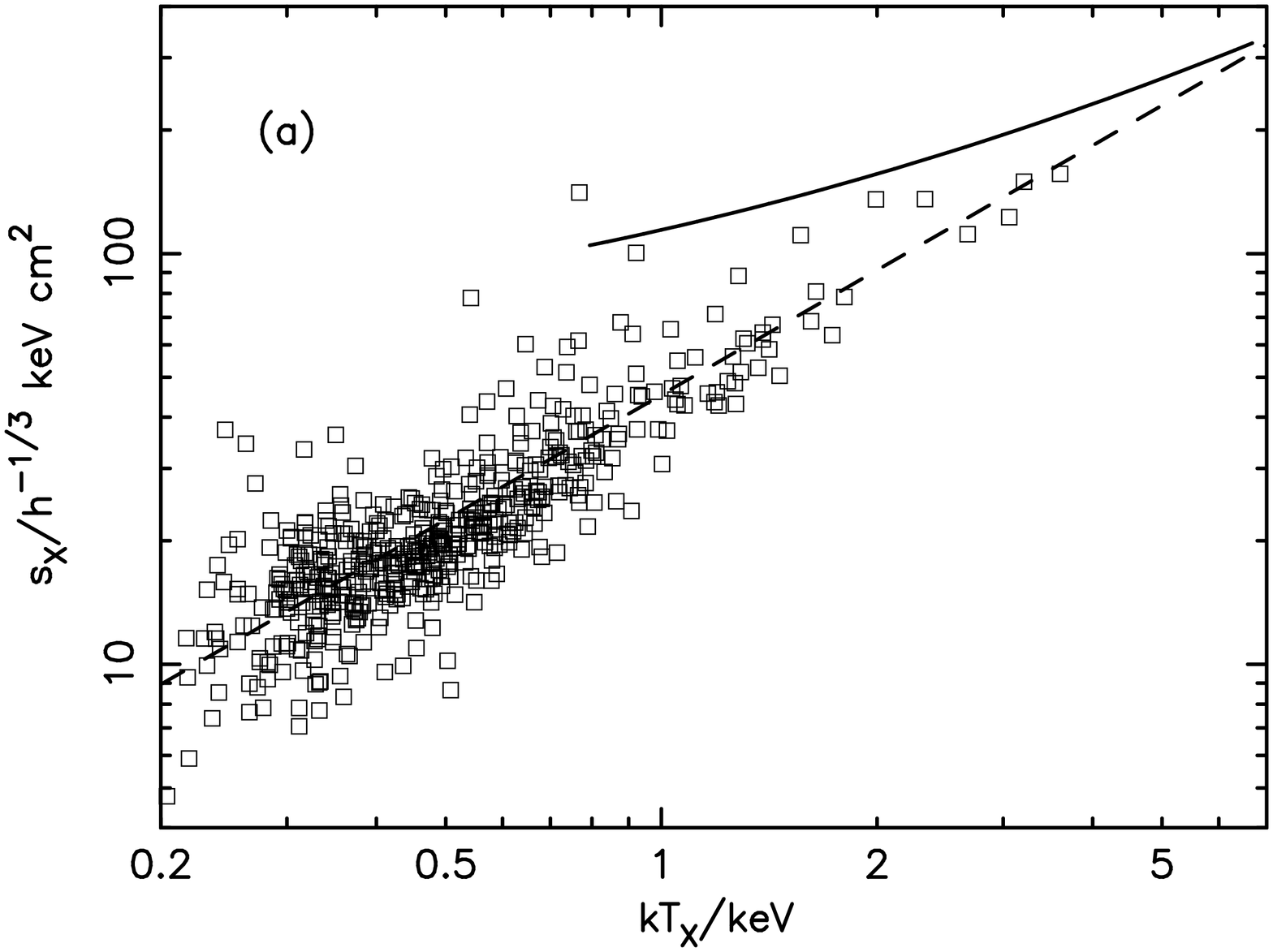,angle=270,width=8.7cm}
\psfig{file=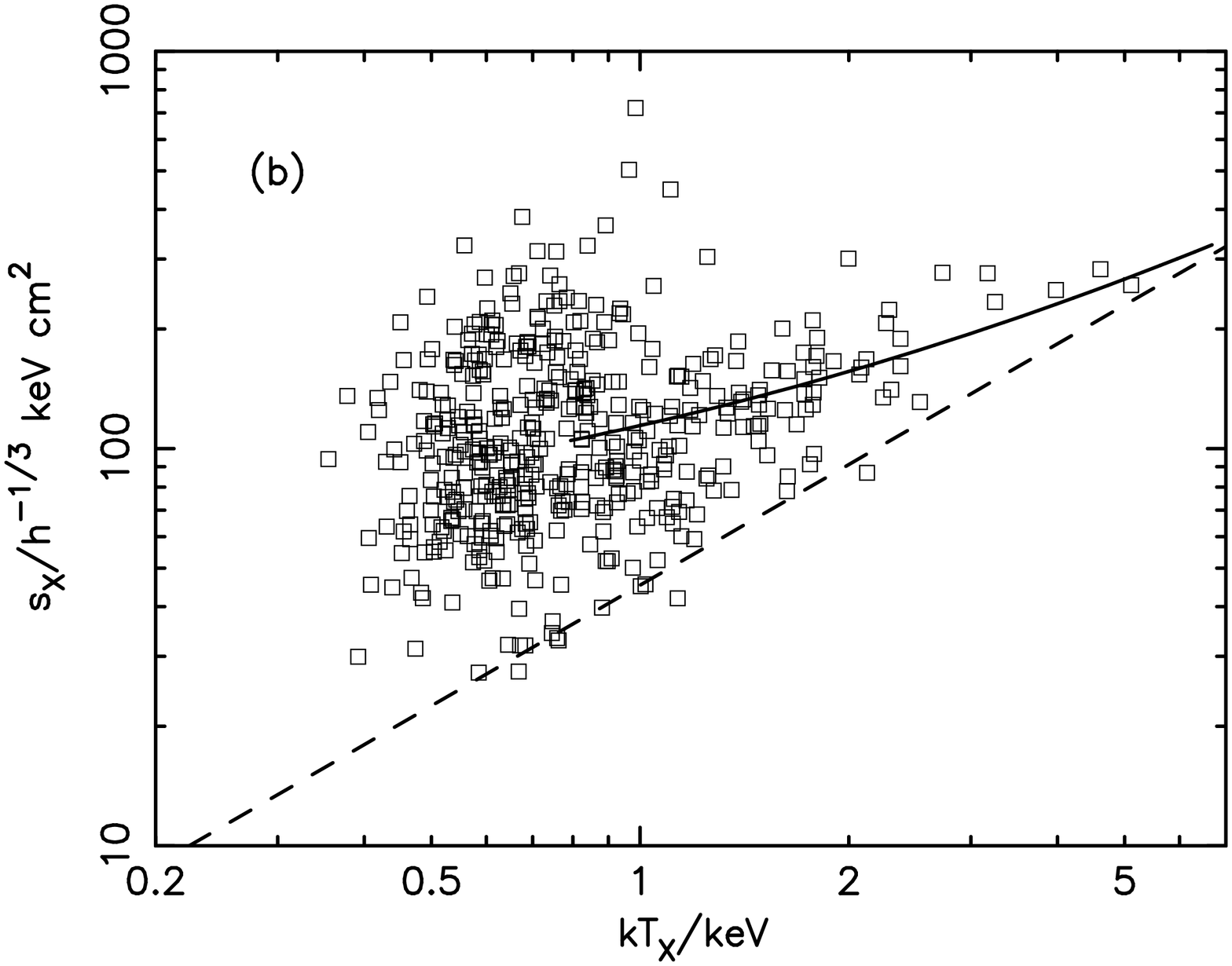,angle=270,width=8.7cm}
\psfig{file=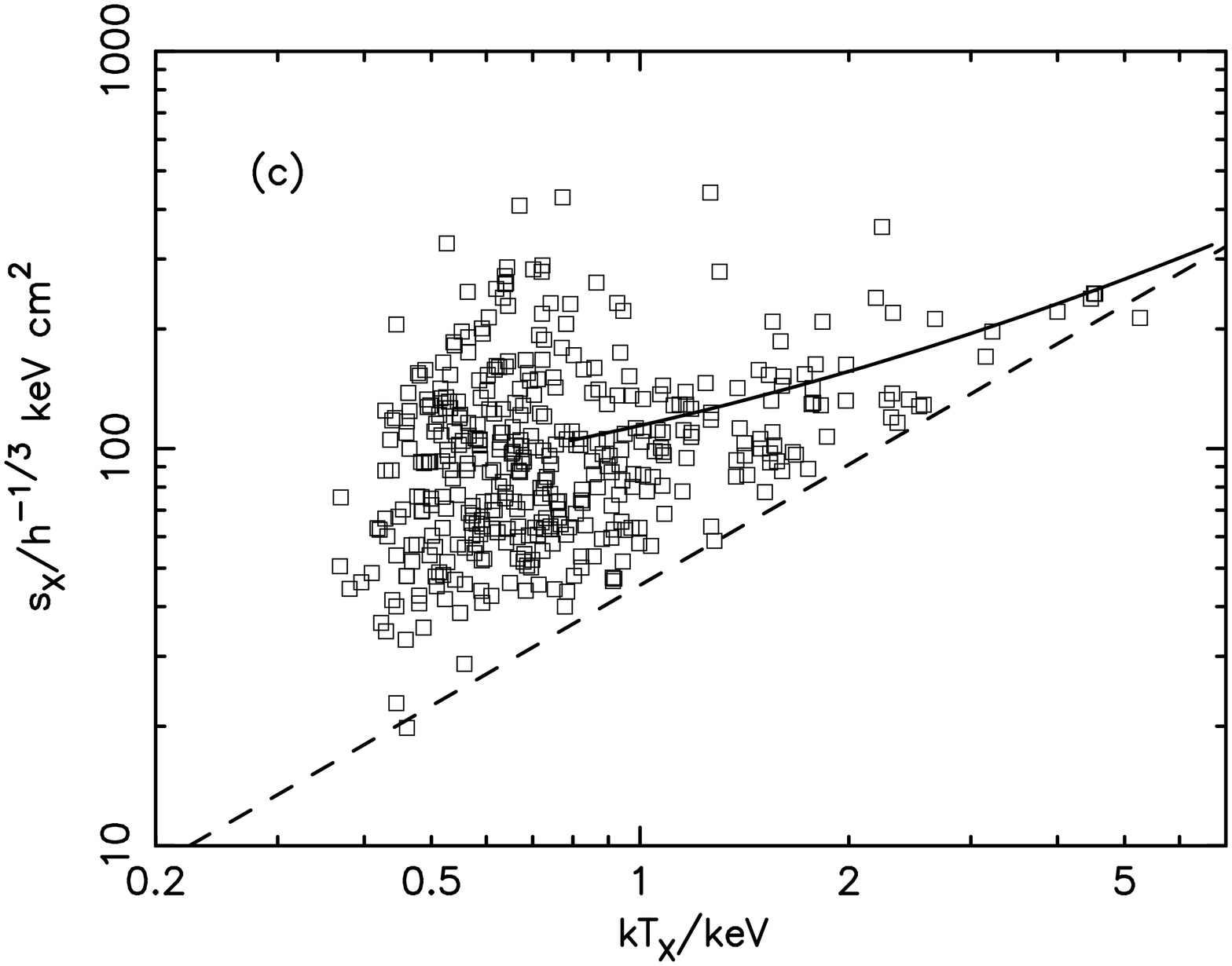,angle=270,width=8.7cm}
\caption{Core entropy at 0.1\,$r_{\rm vir}$ versus
X-ray temperature for clusters in the (a) {\it Non-radiative}, (b)
{\it Radiative}, and (c) {\it Preheating} simulations.  The straight
dashed line is the self-similar relationship from the simulations of
Eke et al.~(1998).  The solid curve is an approximate fit to the
Ponman et al. (1999) data-points from their Fig.~2.}
\label{fig:ecore}
\end{figure}

The core entropy of the clusters in the three simulations are plotted
against their X-ray temperatures in Fig.~\ref{fig:ecore}.  Also shown
is an approximate fit to the observed relation from 
Ponman et al.~(1999, solid curves) and
the prediction from the non-radiative numerical simulations  of Eke,
Navarro \& Frenk~(1998, dashed lines).
In the {\it Non-radiative run} (panel a), the core entropy follows the
self-similar relation well, albeit with large scatter.  On the other
hand, the core entropy of clusters in the {\it Radiative} and {\it Preheating}
simulations (panels b and c, respectively) is in good agreement with
the observations (the solid curves). In the {\it Preheating}
simulation, the excess entropy relative to the self-similar prediction is due 
to the increase in energy of the gas at $z=4$ (given that radiative losses
in this simulation are neglegible, the excess entropy is conserved).
The agreement of the {\it Radiative} simulation with observations,
however, is somewhat fortuitous: in the absense of some form of feedback
machanism, the amount of gas that cools to low temperatures is
determined only by the resolution of the simulation and the imposed
metallicity of the gas.

\subsection{Entropy and temperature profiles}
\label{sec:etprof}

\begin{figure}
\psfig{file=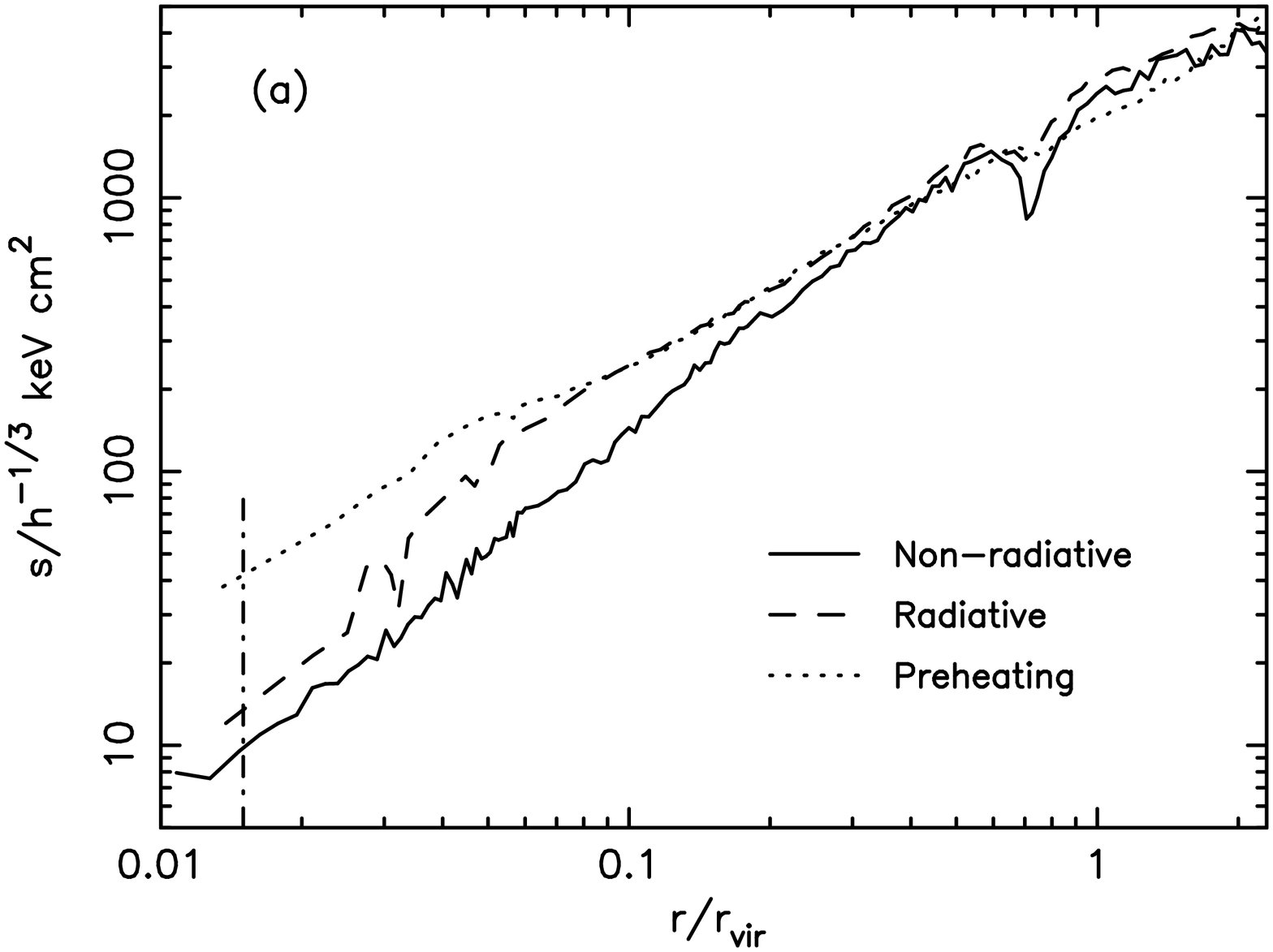,angle=270,width=8.5cm}
\psfig{file=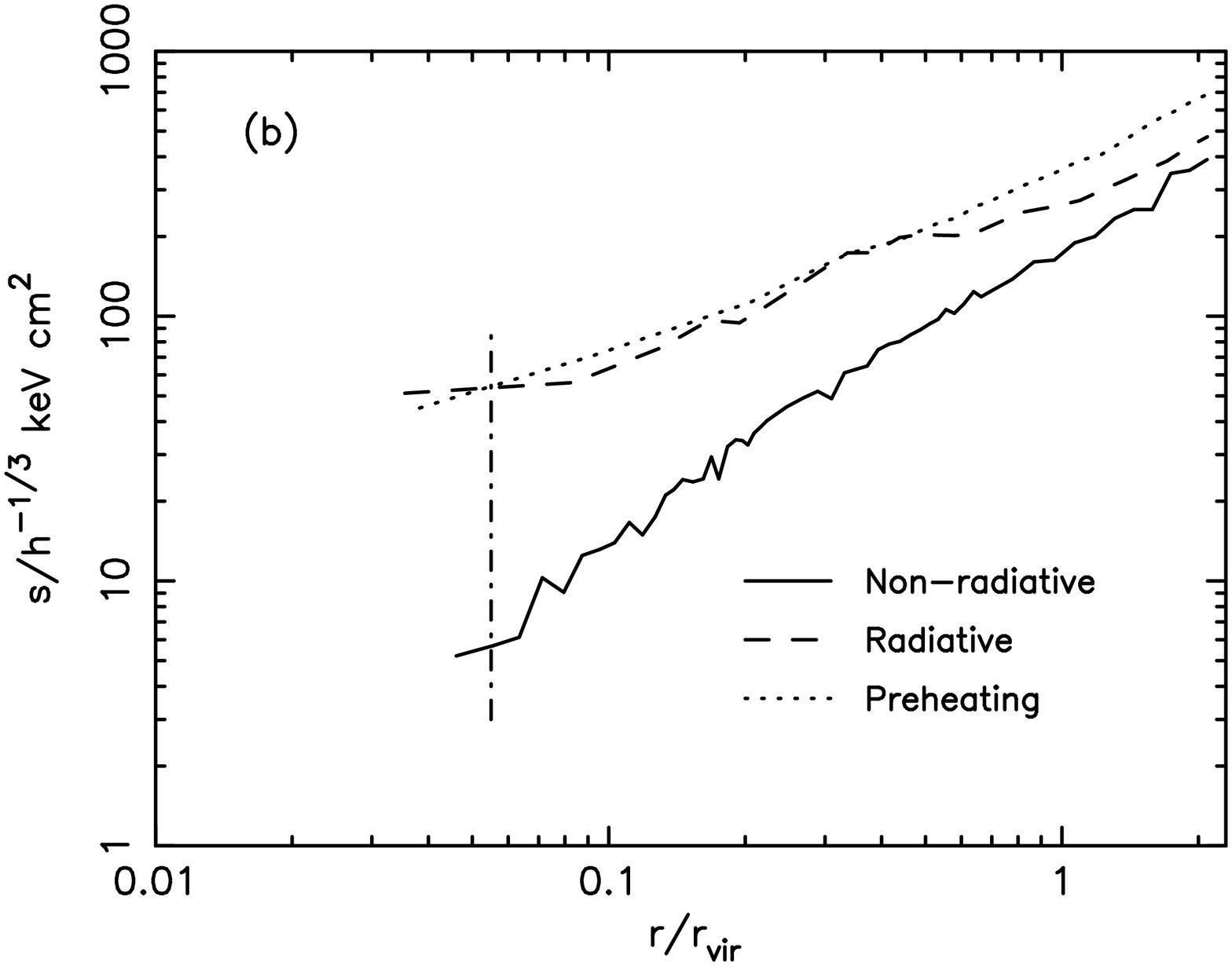,angle=270,width=8.5cm}
\caption{The entropy profiles in each of the three simulations of (a)
the third-most-massive and (b) the 10 least-massive clusters that did
not show significant substructure.  The long dash-dotted line
indicates the gravitational softening length.  Bins were chosen so as
to average over at least 32 particles near the cluster core, with more
further out.}
\label{fig:eprof}
\end{figure}

Fig.~\ref{fig:eprof} shows entropy profiles for high- and low-mass
clusters from each simulation.  The entropy is defined as in
equation~(\ref{eq:entropy}) except that we use the mass-weighted
rather than the emission-weighted temperature.  The high-mass cluster
is the third-largest in the box and was chosen because the two
most-massive clusters show significant amounts of substructure within
their virial radii; even for this cluster, however, there is an obvious
subclump at about 0.7\,$r_{\rm vir}$ that significantly distorts the
entropy profile in the {\it Non-radiative} simulation.  The low-mass
profiles were constructed by averaging the profiles of the 10
lowest-mass clusters in each simulation that did not show significant
substructure---only about half the clusters were deemed acceptable.
Thus neither panel is representative of a typical cluster in our
simulations as most have much more substructure than is visible here.

First note that the entropy profiles of the three runs roughly match
for the massive cluster at radii greater than 0.3\,$r_{\rm vir}$.  The
differences are due to different amounts of cool gas in subclumps that
permeate the outer regions of all clusters.  Within this radius, the
{\it Radiative} and {\it Preheating} runs have significantly higher
entropy than the {\it Non-radiative} one.
For the {\it Preheating} run, this excess entropy arises from the
energy injection at $z=4$ and is subsequently almost unchanged, except for 
a slight decrease within 0.04\,$r_{\rm vir}$, where radiative
cooling has had a small effect.
The entropy increase in the {\it Radiative} simulation, although it
follows that in the {\it Preheating} run closely, comes about through
an entirely different mechanism, namely inflow of high-entropy gas to
replace low-entropy gas lost via cooling. That the two agree is
coincidental except in as much as we have chosen the simulation
parameters such that the two runs have similar X-ray properties.

For the low-mass clusters the behaviour is similar, except that the
excess entropy extends out to a larger fraction of the virial radius.  
The {\it Radiative} and {\it Non-radiative} runs have converged by
$r=2\,r_{\rm vir}$ since cooling is efficient only in the centre
of the cluster. In the {\it Preheating} run, however, the excess entropy 
is still apparent beyond $2\,r_{\rm vir}$ as the entropy was increased 
for all of the gas.

For all clusters in the {\it Preheating} simulation, the excess entropy 
is much less than expected for gas at the mean density of the box at the 
time of preheating (about 770\,$h^{-1/3}$\,keV\,cm$^2$). This is because 
the intracluster gas at $z=0$ was already in overdense structures by $z=4$. 
For example, the average density of the gas in the 10 least massive 
clusters is about 20 times the mean gas density at $z=4$. Thus, 
the additional energy injected into
the gas only increased the entropy by approximately 
100\,$h^{-1/3}$\,keV\,cm$^2$, as is evident from Fig.~\ref{fig:eprof}.

\begin{figure}
\psfig{file=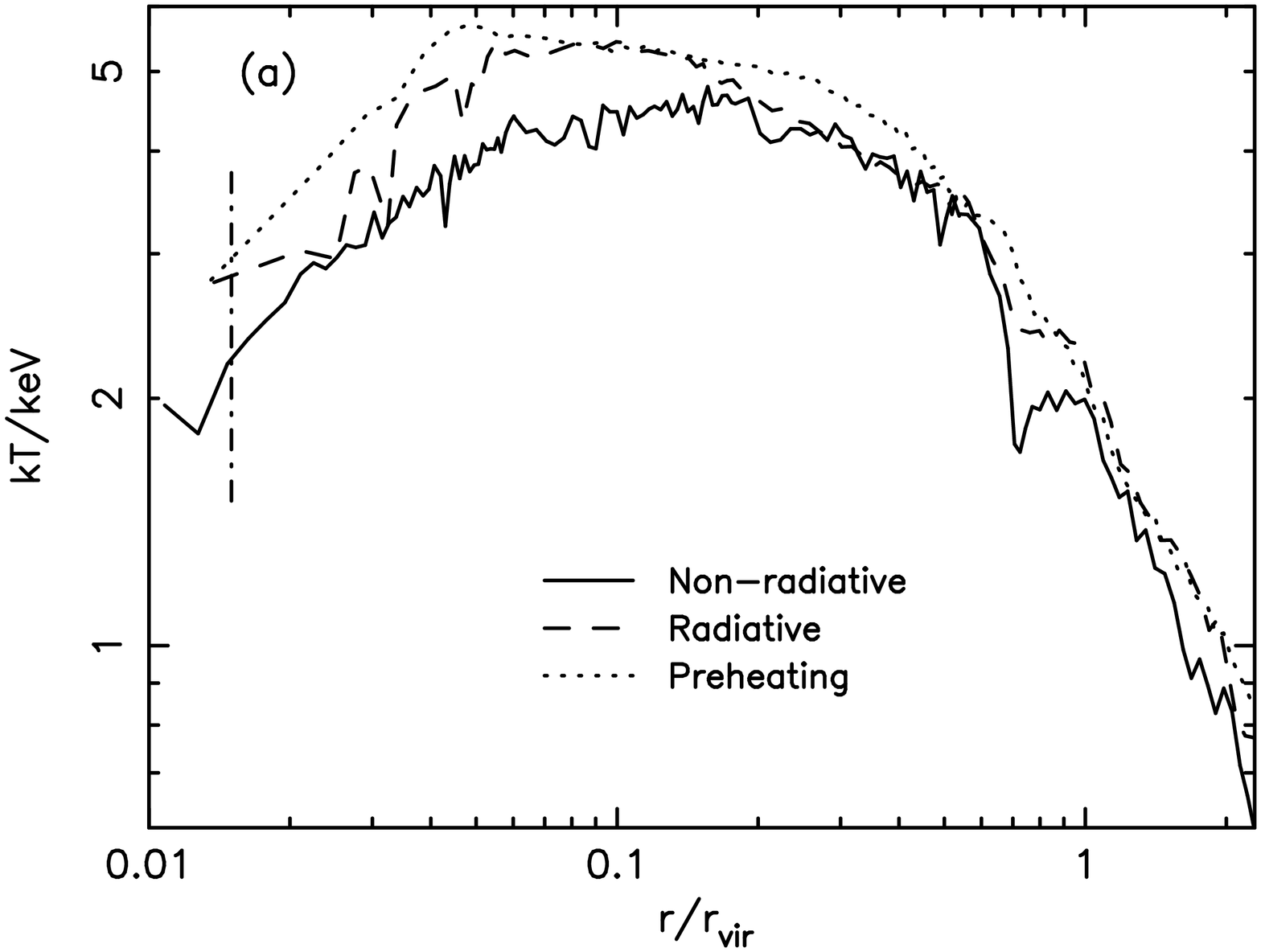,angle=270,width=8.5cm}
\psfig{file=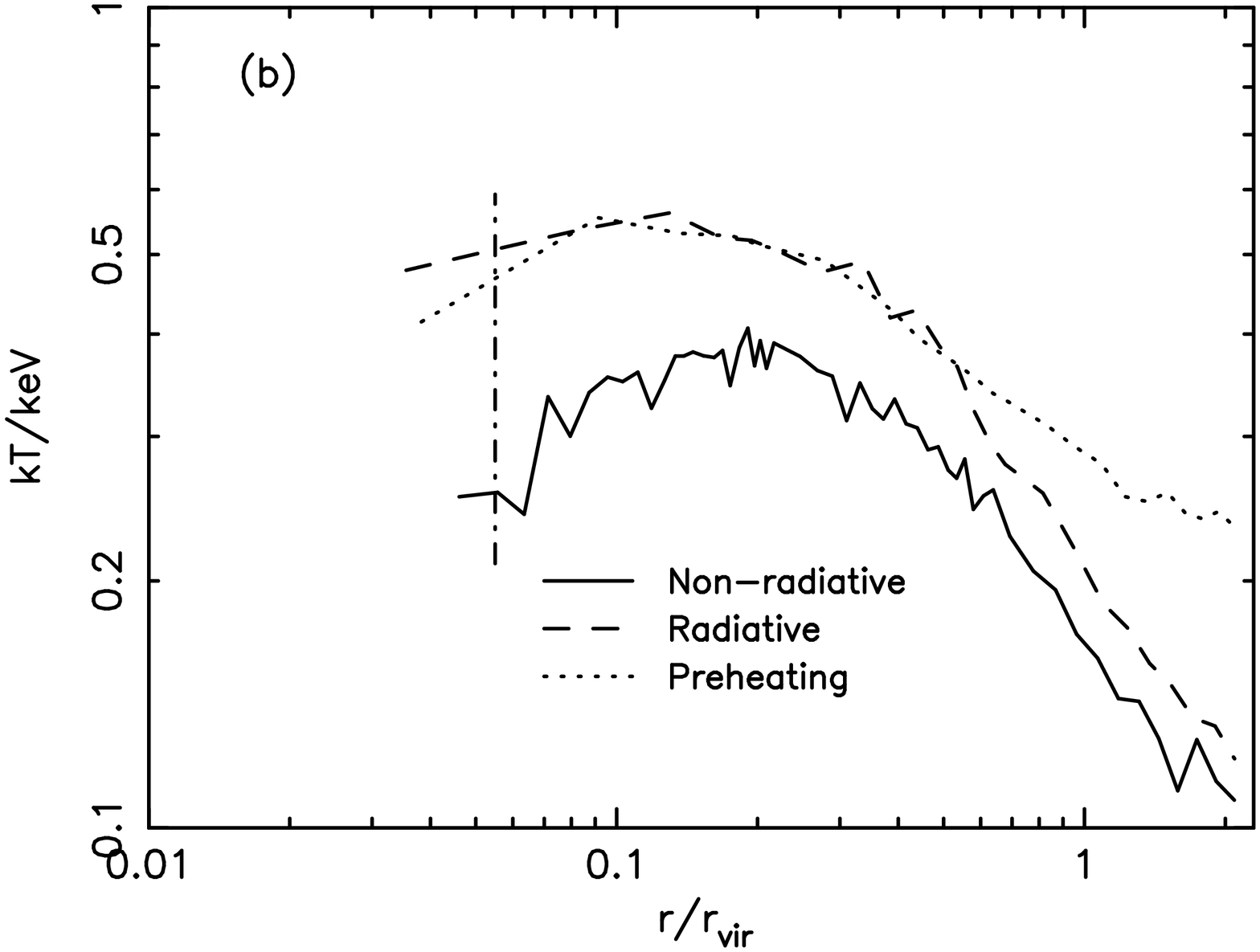,angle=270,width=8.5cm}
\caption{The mass-weighted temperature profiles of (a) the third
most-massive and (b) the 10 least-massive clusters that did 
not show significant substructure.  The long dash-dotted line
indicates the gravitational softening length.  Bins were chosen so as
to average over at least 32 particles near the cluster core, with more
further out.}
\label{fig:tprof}
\end{figure}

Fig.~\ref{fig:tprof} shows mass-weighted temperature profiles for
the same clusters as in Fig.~\ref{fig:eprof}.
Emission-weighted temperatures are similar to these in the inner
parts of clusters but decline more rapidly in their outer regions
dues to the presence of dense, cool subclumps.
Note that the excess temperature in the intracluster medium in the
{\it Preheating} simulation is much less than 1\,keV/k, except near
the cores of the largest clusters. Thus, the gas must have cooled
adiabatically, flowing out of clusters and reducing its density, since
the time it was heated.  By contrast, the gas in the {\it Radiative}
run has flowed inwards, raising its temperature by adiabatic compression.

In their inner regions, the clusters are approximately isothermal.
However, the gas temperature declines rapidly at large radii,
typically dropping by a factor of 2--3 from its peak out to the virial
radius, in agreement with results from previous simulations
(e.g. Frenk et al. 1999).  Recent observational evidence finds no
evidence of departures from isothermality of X-ray clusters (Irwin,
Bregman \& Evrard 1999; White 2000; see, however, Markevitch
et~al.~1998) but only probe out to about 0.4\,$r_{\rm vir}$, just
where the temperature profiles steepen.  If these observations can be
extended out to larger radii then this would be a strong test of the
models.

\subsection{Surface Brightness profiles}
\label{sec:sbprof}

\begin{figure}
\psfig{file=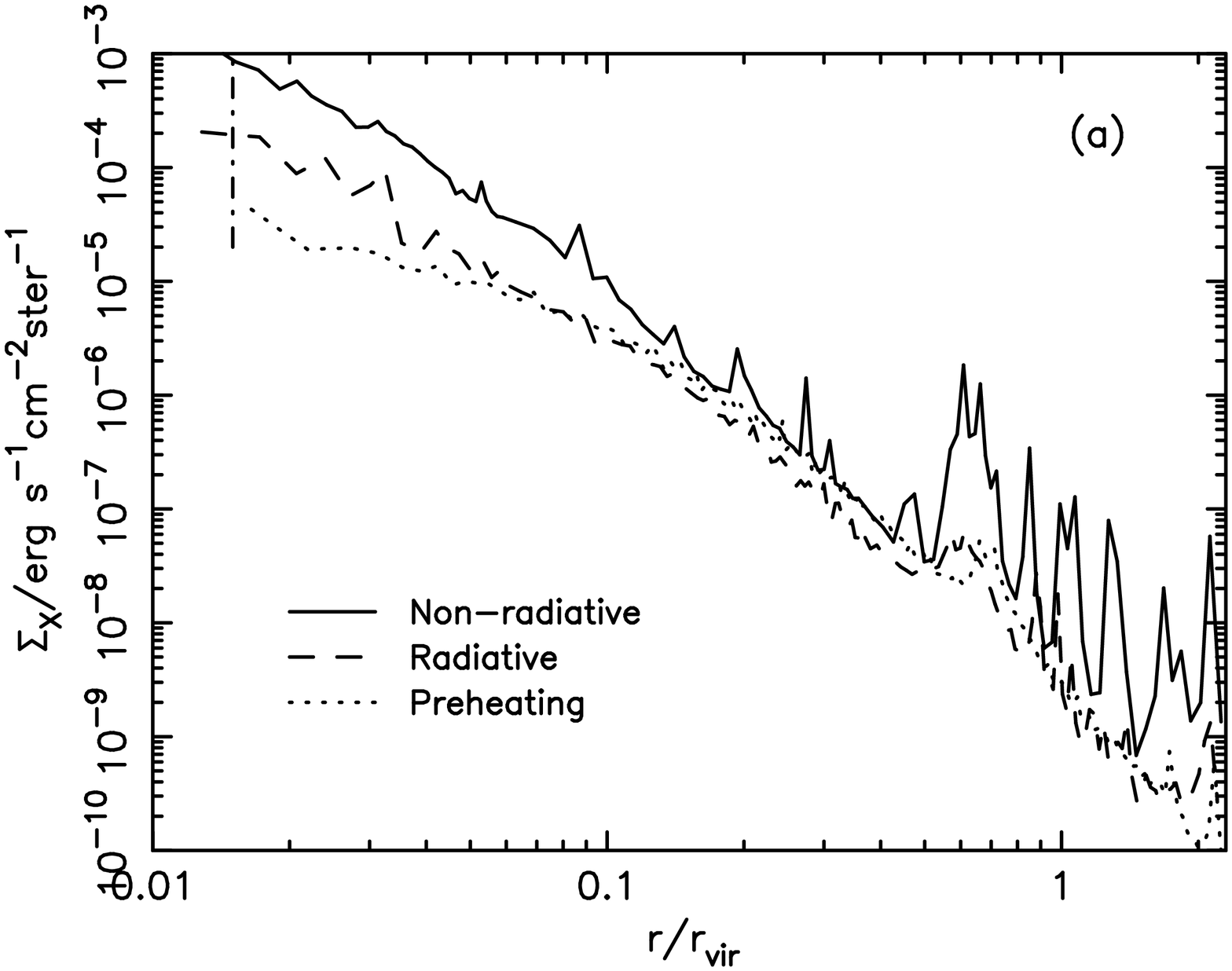,angle=270,width=8.5cm}
\psfig{file=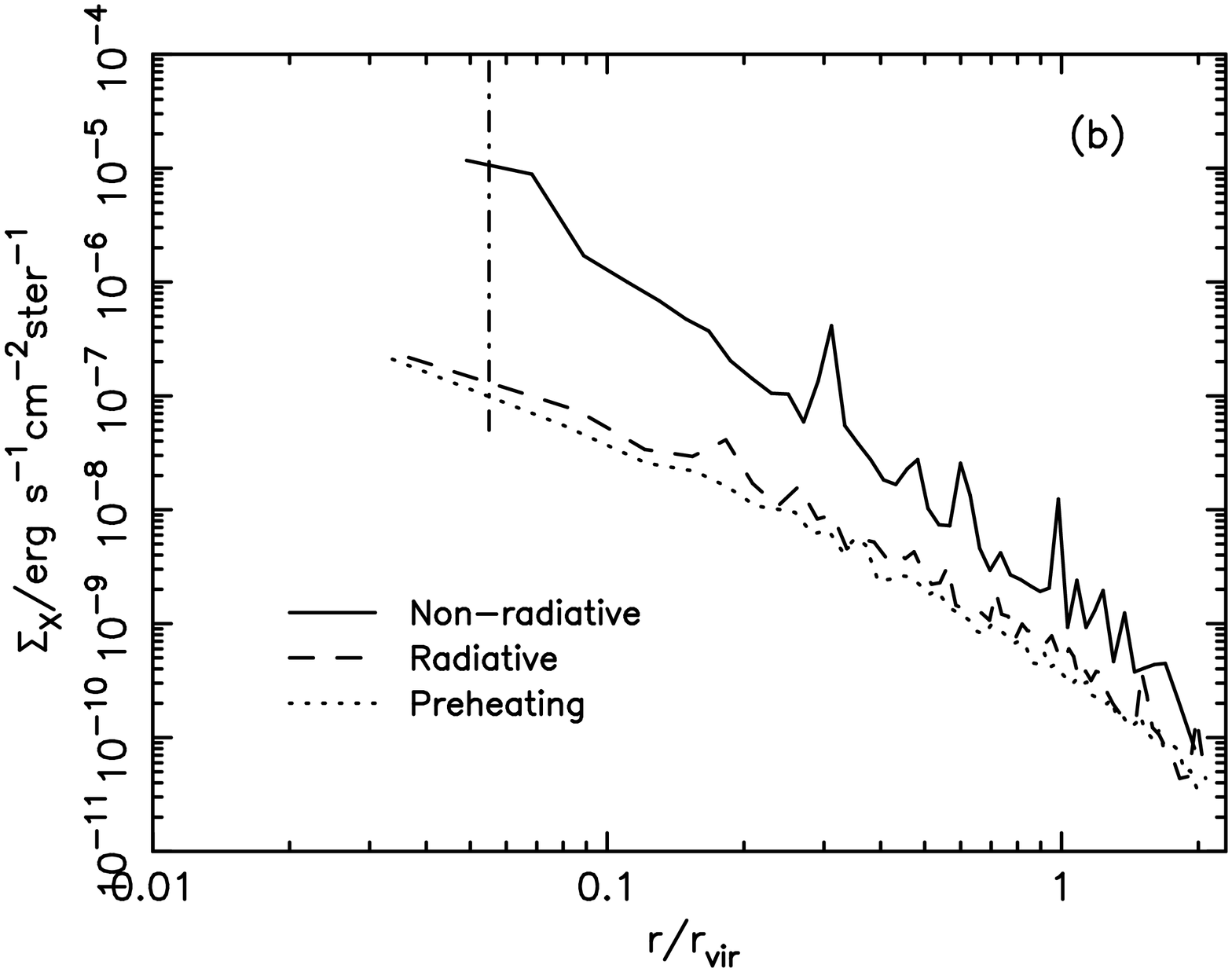,angle=270,width=8.5cm}
\caption{The averaged X-ray surface brightness profiles in the 
0.3--1.5\,keV band in each of the three simulations of (a) the
third-most-massive and (b) the 10 least-massive clusters that did not
show significant substructure.  The long dash-dotted line indicates
the gravitational softening length.  Bins were chosen so as to average
over at least 64 particles near the cluster core, with more further out.}
\label{fig:sbprof}
\end{figure}

X-ray surface brightness profiles for the third-most-massive and an
averaged low-mass cluster are shown in Fig.~\ref{fig:sbprof}.  Note,
once again, the significant substructure that is present even in these
profiles that have been specially selected to be as smooth as possible.
This substructure is particularly prevalent in the X-ray
surface brightness because of its weighting with the square of the gas
density.

The total mass in the intracluster medium within the virial radius of
the massive cluster in the {\it Preheating} and {\it Non-radiative}
simulations is roughly equal, yet the {\it Non-radiative} simulation
has a higher surface brightness at all radii---this is because of
emission from high density gas in subclumps that is heated and
expelled in the {\it Preheating} run.

The isothermal-$\beta$ model (Cavaliere \& Fusco-Femiano 1976) is 
commonly used to estimate cluster masses. The model assumes a halo
to be a spherical, isothermal sphere. The surface brightness
profile of an isothermal gas in hydrostatic equilibrium within such a
halo is
\begin{equation}
\Sigma_X(R)=\Sigma_X(0)\big(1+(R/R_{\rm c})^2\big)^{1/2-3\beta_{\rm fit}}
\label{eq:isobeta}
\end{equation}
where $R_c$ is the X-ray core radius.

It is not at all obvious that equation~(\ref{eq:isobeta}) provides a
good description of the surface brightness profiles in real clusters.
These often fail to flatten in their inner regions as much as is
predicted by equation~(\ref{eq:isobeta}), so the central regions are
then omitted from the fitting and the excess emission over the
isothermal-$\beta$ model is attributed to a cooling flow.  Furthermore,
the outer radius to which one can measure the surface brightness
accurately is only a small fraction of the virial radius and depends
upon the size and distance of the cluster.

We now estimate the values of $\beta_{\rm fit}$ for our simulated
clusters.  Then later, in Section~\ref{sec:tm}, we will test the
reliability of cluster masses deduced using the isothermal-$\beta$ model.  An accurate comparison with X-ray observations would
require us to produce mock X-ray data from our simulations, with
appropriate particle backgrounds, etc., and then to run these through
the X-ray analysis software.  We do not yet have the tools to do this
so instead adopt a simpler procedure that is robust even in the
presence of substructure.
As seen in Fig.~\ref{fig:sbprof}, the surface brightness profiles
of the simulated clusters fluctuate, even well within the virial radius.
In order to smooth the profiles, we integrate them inwards to obtain 
a cumulative luminosity profile.  Using the isothermal-$\beta$ model,
this can be expressed as
\begin{equation}
L_X(>R)= L_X(0)\big(1+(R/R_{\rm c})^2\big)^{3/2-3\beta_{\rm fit}},
\label{eq:lproj}
\end{equation} 
where $R$ denotes a 2-dimensional, projected radius.
We need a statistic that is insensitive to
emission from the outer parts of clusters (where most of the
substructure originates) and from a possible central cooling flow.
One such is
\begin{equation}
{\cal L} =\frac{L_X(>R_{\rm mid})-L_X(>R_{\rm out})}
	{L_X(>R_{\rm in})-L_X(>R_{\rm out})},
\label{eq:lprojrat}
\end{equation}
which depends only on the flux between $R_{\rm in}$ and $R_{\rm out}$,
where $R_{\rm in}<R_{\rm mid}<R_{\rm out}$.

In our fits, we fix $R_{\rm out}=0.4\,r_{\rm 200}$ which is
approximately equal to the outer extent of observed X-ray
surface brightness profiles.  We also fix $R_{\rm in}=R_{\rm c}=R_{\rm
KS}$ where $R_{\rm KS}$ is based on the model of Komatsu \&
Seljak (2001). They determined a relationship between the core radius of
the X-ray emission and the concentration parameter of the dark matter
density profile (Navarro et al. 1997), 
\begin{equation}
R_{\rm KS}\approx0.4\,r_{500}/c,
\label{eq:ks}
\end{equation}
where 
\begin{equation}
c=6\left(\frac{M_{vir}}{10^{14} \hMsol}\right)^{-1/5}.
\end{equation}
(We note that that Komatsu \& Seljak inadvertently put the
factor of 0.4 above into the denominator rather than the numerator in
Equation~\ref{eq:ks};
the expression as we give it is compatible with their Fig.~13.)

\begin{figure}
\psfig{file=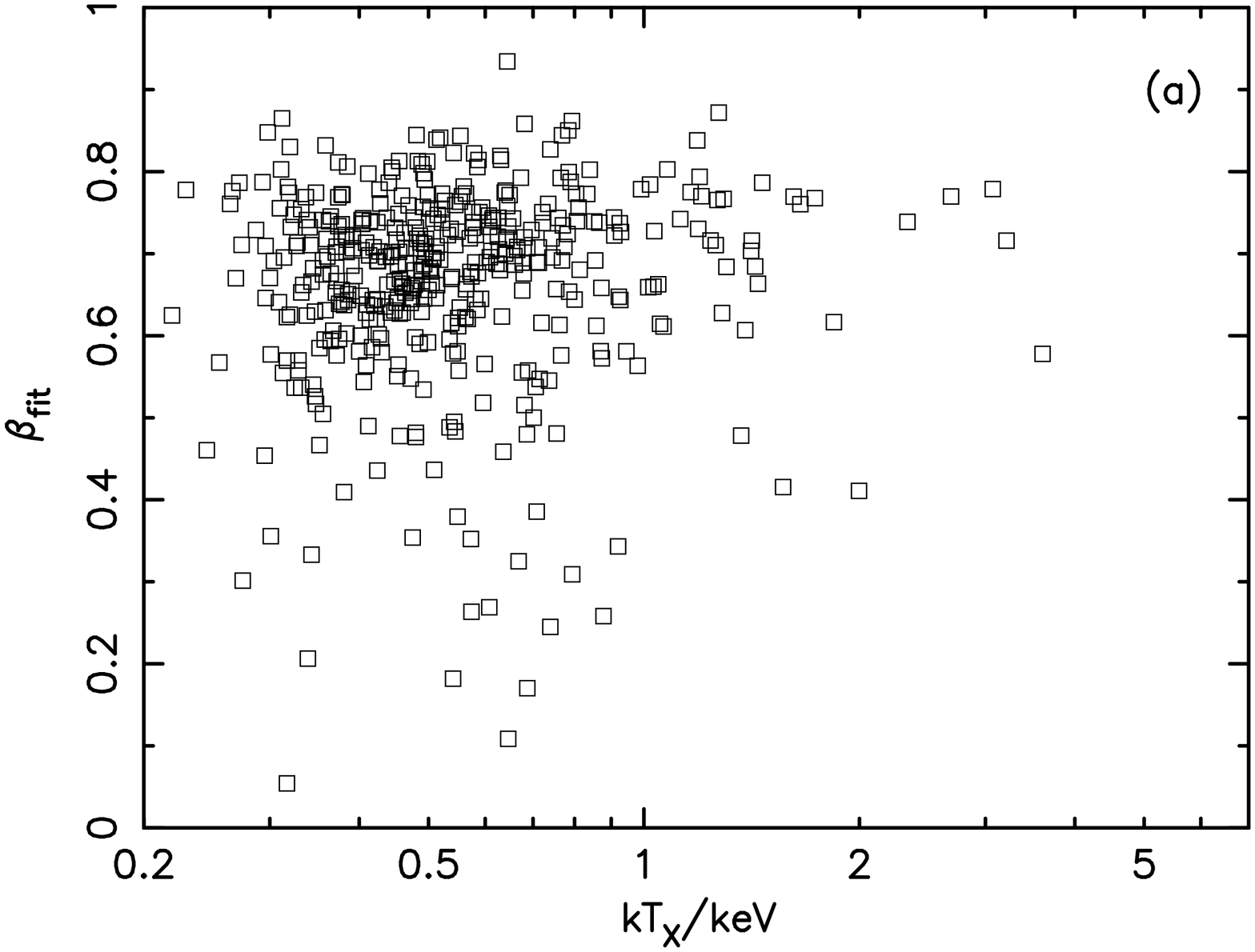,angle=270,width=8.5cm}
\psfig{file=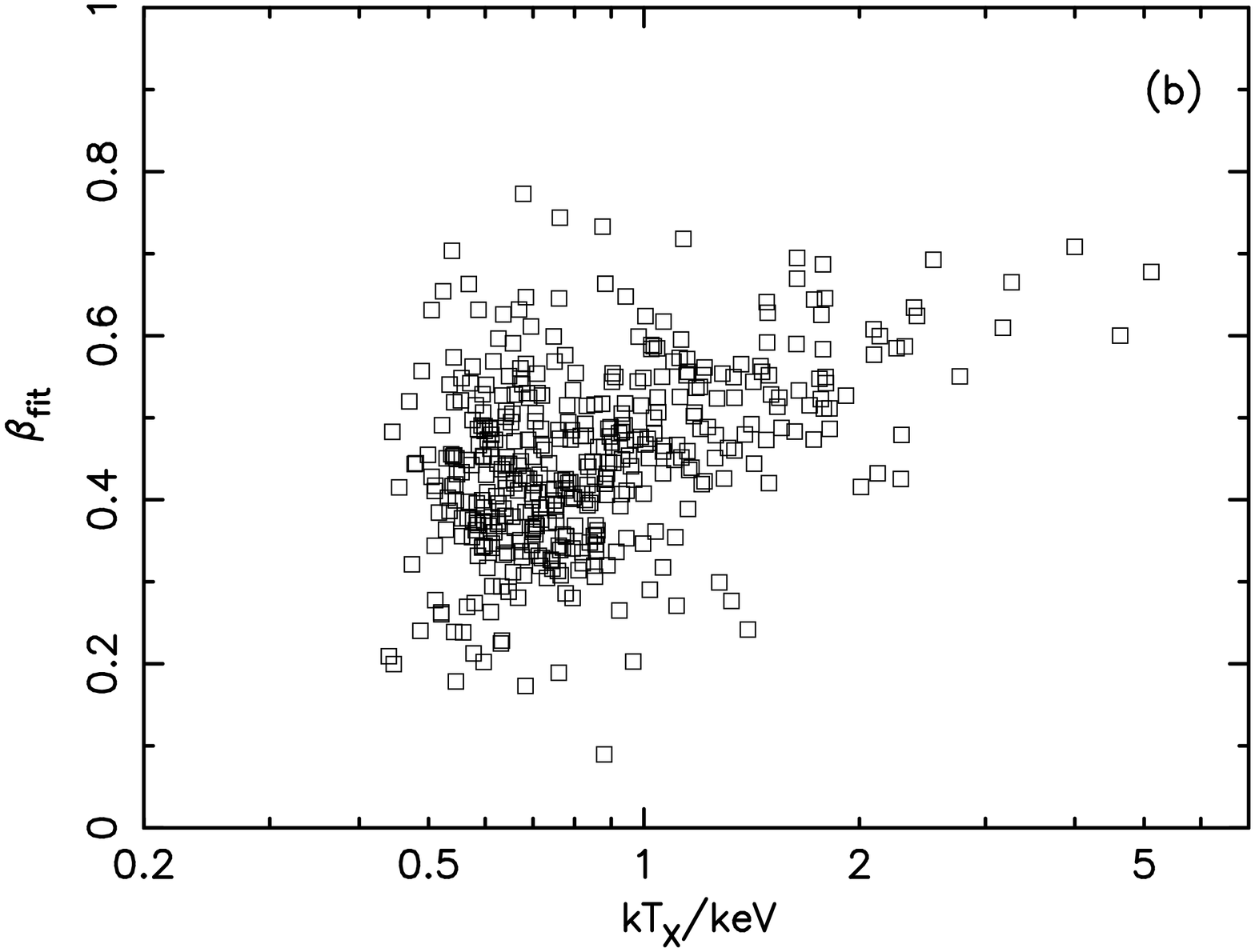,angle=270,width=8.5cm}
\psfig{file=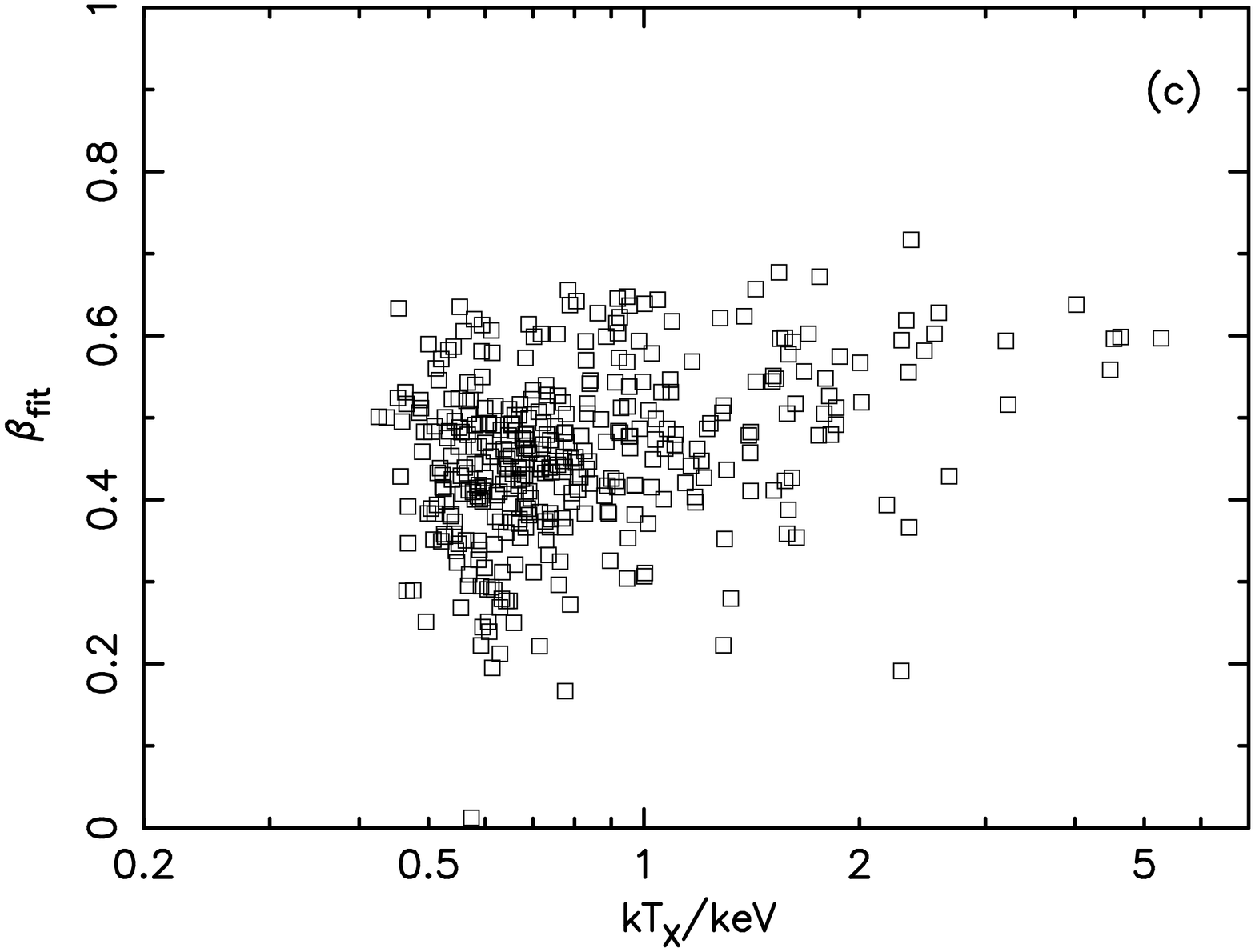,angle=270,width=8.5cm}
\caption{The surface brightness fitting parameter, $\beta_{\rm fit}$, as
a function of temperature in the 0.3--1.5\,keV band, for clusters
drawn from the (a) {\it Non-radiative}, (b) {\it Radiative} and
(c) {\it Preheating} simulations.}
\label{fig:betafit}
\end{figure}

Finally, we take $R_{\rm mid}=0.15\,r_{200}$ and then solve
Equations~\ref{eq:lproj} and \ref{eq:lprojrat} for $\beta_{\rm fit}$
with the results shown in Figure~\ref{fig:betafit}.

The first thing to note is the scatter in the calculated values of
$\beta_{\rm fit}$. This is because individual surface brightness
profiles are very noisy due to the presence of substructure.
Secondly, there is a trend for the mean value of $\beta_{\rm fit}$ to
decrease with decreasing temperature in the {\it Radiative} and {\it
Preheating} simulations with the value of approximately 2/3 in large
clusters. The trend is much less apparent in the {\it Non-radiative} model
where most clusters throughout the temperature range have consistently
high values of $\beta_{\rm fit}$.  Komatsu \& Seljak (2001) propose
that such a trend may arise from observational bias due to the fact
that one observes out to a larger fraction of the virial radius in
higher-mass clusters; that is not the explanation here as we have fixed
$R_{\rm out}=0.4\,R_{200}$ (we tried taking a variable outer radius
as described by Komatsu \& Seljak but found that it made little
difference to our results).  Rather, the difference in $\beta_{\rm
fit}$ values between the {\it Non-radiative} and the other two runs is
a real one, reflecting the raised entropy and hence the reduced
density in the inner regions of low-mass clusters in the {\it
Radiative} and {\it Preheating} runs.

If the data were of sufficiently high quality, one could evaluate
${\cal L}$ at two different choices of $R_{\rm mid}$ and then solve for
both $R_c$ and $\beta_{\rm fit}$.  In practice, there
is a strong correlation between these two parameters and the solutions
sometimes return unacceptable large values of $R_{\rm c}$ (greater than
$R_{\rm out}$) and correspondingly large values of $\beta_{\rm fit}$.
It is for this reason that we fix the core radius before fitting.
Fortunately, $\beta_{\rm fit}$ is insensitive to the precise choice of
core radius: we have checked that doubling the value of the core radius,
i.e.~setting $R_{\rm c}=2\,R_{\rm KS}$, does not alter the results
significantly.  We have also taken the more conventional approach
of fitting functions of the form of Equation~\ref{eq:isobeta} directly
to the surface brightness profiles.  This again gives similar results
but with larger scatter.  This is because the surface brightness
fitting is more sensitive to the presence of substructure at the outer
edge of the fit than is the method that we describe above.

The measurement of the $\beta_{\rm fit}$ parameter as presented above
is far removed from the usual observational practice. Nevertheless,
we shall see in Section~\ref{sec:tm} that the total mass within $r_{200}$
when determined using the parameter is in good agreement with
observations.

\subsection{Cooling flows}
\label{sec:coolflow}

There is currently considerable confusion over the observational
evidence for cooling flows in clusters of galaxies.  Recent results
show both a deficit of soft X-ray emission lines (e.g.~Peterson
et~al.~2002) but also evidence for low-temperature gas at
approximately 10$^5$K (Oegerle et~al.~2001) and even large amounts of
molecular gas (Edge 2001).  Our simulations do not have the resolution
to investigate cooling flows in detail.  Here we simply wish to test
whether the standard method of calculating mass-deposition rates gives
the correct answer.

In this Section, we calculate mass-deposition rates using
two methods: (i) by directly measuring the rate at which gas cools
out of the hot intracluster medium to temperatures below 10$^5$K
and (ii)
by measuring the emission from within the cooling radius, defined
as the radius within which the gas has a mean cooling time
less than 6 Gyr.

\begin{figure}
\psfig{file=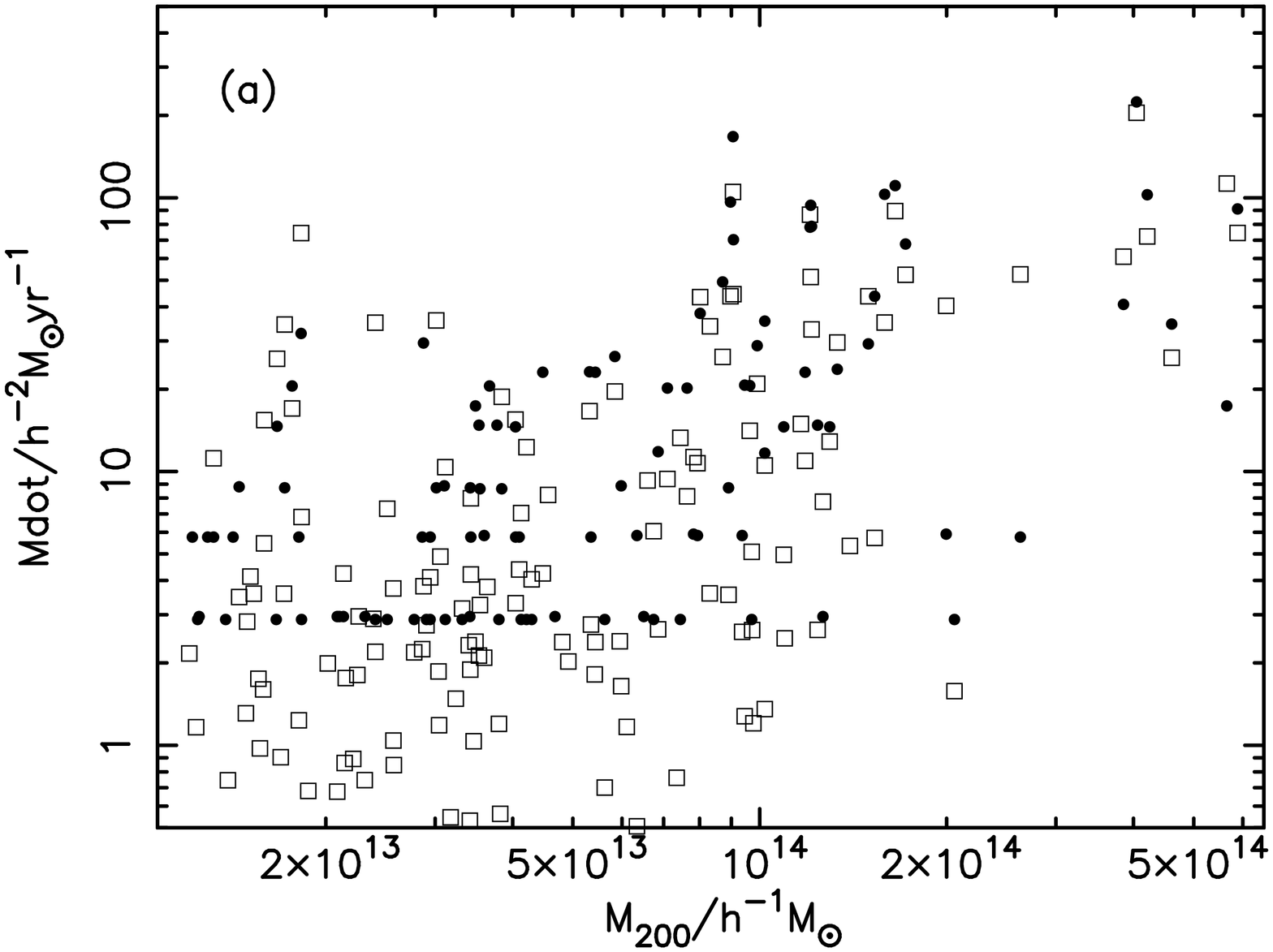,angle=270,width=8.5cm}
\psfig{file=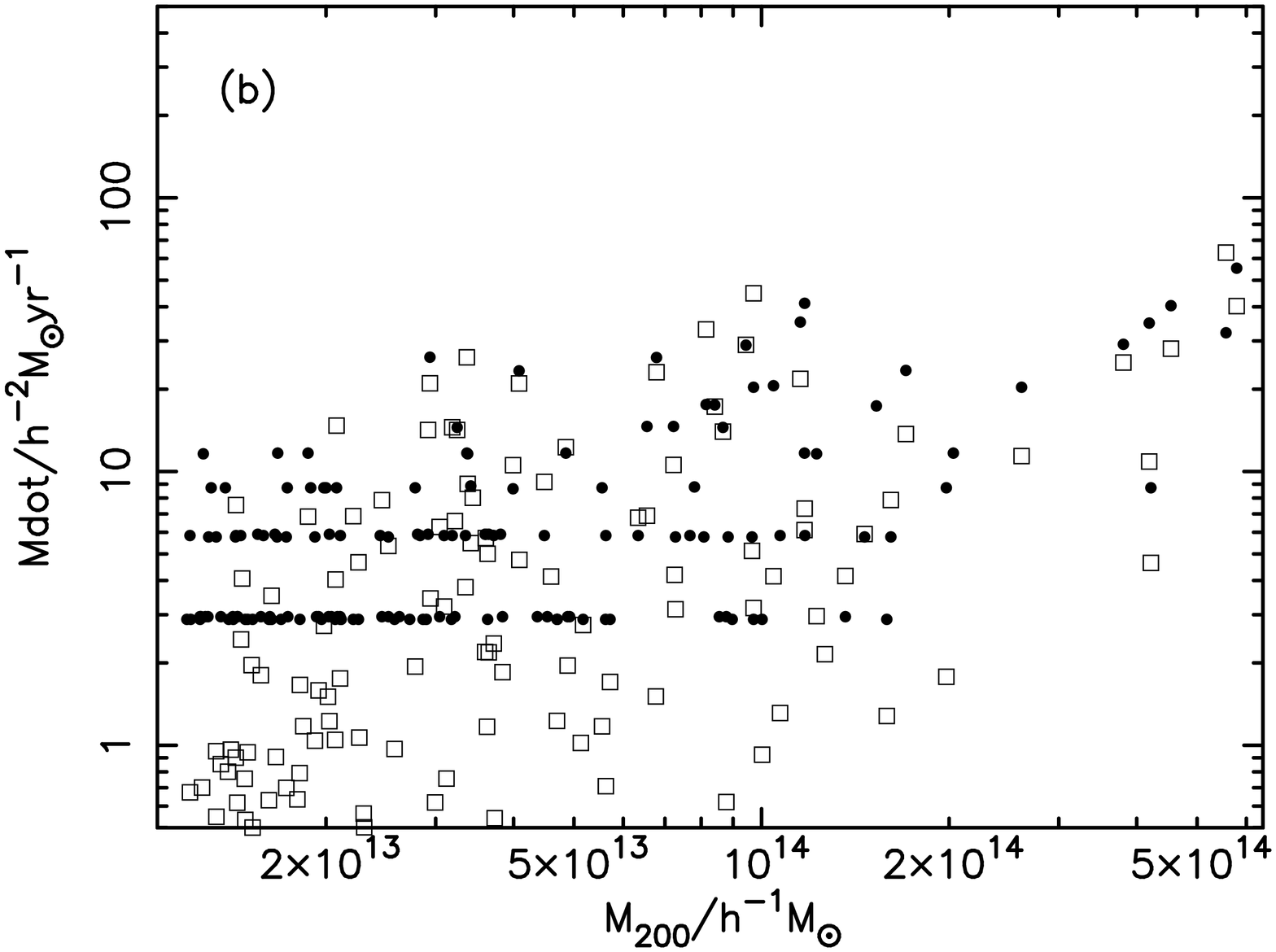,angle=270,width=8.5cm}
\caption{The mass-depostion rates for clusters drawn from (a) the {\it
Radiative} and (b) the {\it Preheating} simulations.  The open squares
show the predicted mass-deposition rate based on the soft-band X-ray
emission from within the cooling radius; the solid circles show the
actual mass-deposition rate averaged over approximately 1 Gyr.}
\label{fig:mdotcmp}
\end{figure}

Fig.~\ref{fig:mdotcmp} illustrates the two measures of the mass
deposition rate for clusters in the {\it Radiative} and 
{\it Preheating} simulations.  The solid circles show the actual mass
deposition rate averaged over the last few output times in our
simulations, corresponding to a time interval of just under 1\,Gyr.
Although the mass is measured in \hMsol, we have plotted it in units
of $h^{-2}$\Msol yr$^{-1}$ to aid comparision with the predicted rate.
The particle resolution limits the measurement of the mass-depostion
rate to be a multiple of approximately 3\,$h^{-2}$\Msol
yr$^{-1}$.  Note that the mass-deposition is extrememly stochastic in
that the mass-deposition rate in the first half of this time interval
typically differs from that in the second half by a factor of more
than two.

The open squares show the predicted mass-deposition rate, defined as
\begin{equation}
\dot{M}_{\rm pred}={L_{X, {\rm rcool}}
\over5kT_{X, {\rm rcool}}/2\mu m_{\rm H}},
\label{eq:mdotpred}
\end{equation}
where $L_{X, {\rm rcool}}$ and $T_{X, {\rm rcool}}$ are the bolometric
luminosity and mean temperature within the cooling radius within which
the gas has a mean cooling time of less than 6\,Gyr.  The temperature
and luminosity here are estimated from emission in the soft band, as
described in equations~(\ref{eq:ktx}) and (\ref{eq:lx}).  Although
there is a large scatter, the predicted and actual mass-deposition
rates roughly agree with no evidence of a systematic bias of one above
the other.  Because of uncertainties in the definition of the
predicted cooling rate, one should not make too much of this
agreement.  Nevertheless, it does eliminate the possibility, within
the context of the models that we simulate, that incorrect
interpretation of the X-ray observations has led to a vast
overestimate of the actual mass-deposition rate in clusters.

\section{Scaling relations}
\subsection{Temperature--Mass relation}
\label{sec:tm}

The most direct way to compare the simulated
temperature--mass\footnote{We use temperature--mass rather than
mass--temperature because we are complete in mass rather than
temperature.} relation between simulated and observed clusters is to
use the thermal (mass-weighted) temperature of the gas within a small
region that is well-observed in X-rays. We have already done this
and presented the results in a short paper (Thomas et~al.~2002), 
in which we compared the simulations described in this paper to 
observations of 5 relaxed clusters using the {\it Chandra} satellite 
(Allen, Schmidt \& Fabian 2001).  In particular, we
compared the normalisation of the temperature--mass relation for matter
within $r_{2500}$ (where $r_\Delta$ is the radius of the sphere that
encloses a mean density of $\Delta$ times the critical density).  The
{\it Non-radiative} simulation agrees with previous simulations of
that kind (e.g.~Mathiesen \& Evrard 2001) in predicting temperatures
that are too low for a given mass, whereas both the {\it Radiative}
and {\it Preheating} simulations reproduce the observations.
Unfortunately, spatially-resolved temperature data are as yet
available only for a few bright clusters and so the overlap in mass
between the observed and simulated clusters is small; nevertheless,
there is no reason to expect that our simulated results should not
extend up to higher temperatures.

A much larger body of data exists for emission-weighted temperatures
of clusters, with poor spatial resolution.  In this case some form of
modelling is required in order to derive the mass.  The
emission-weighted temperature, being dominated by the high
surface brightness, central regions of the clusters, does not change
very much with $r_\Delta$, but the mass does.  Generally, one wants to
choose as small a radius as possible in order to minimise the
extrapolation outside the region that is well-observed in X-rays.  On
the other hand, theoretical predictions are for the mass within the
virial radius ($\Delta=111$ in this cosmology).  In this paper, we
compromise and use $\Delta=200$ as this is the overdensity used in two
observational papers that we wish to compare with: Horner, Mushotzky
\& Scharf (1999) and Xu, Jin \& Wu (2001).  A third, Finoguenov,
Reiprich \& B\"ohringer (2001) uses $\Delta=500$ but is easily
extrapolated to $\Delta=200$ (using $M\propto\Delta^{-1/2}$ for the
isothermal-$\beta$ model at radii much greater than the core radius).

\begin{figure}
\psfig{file=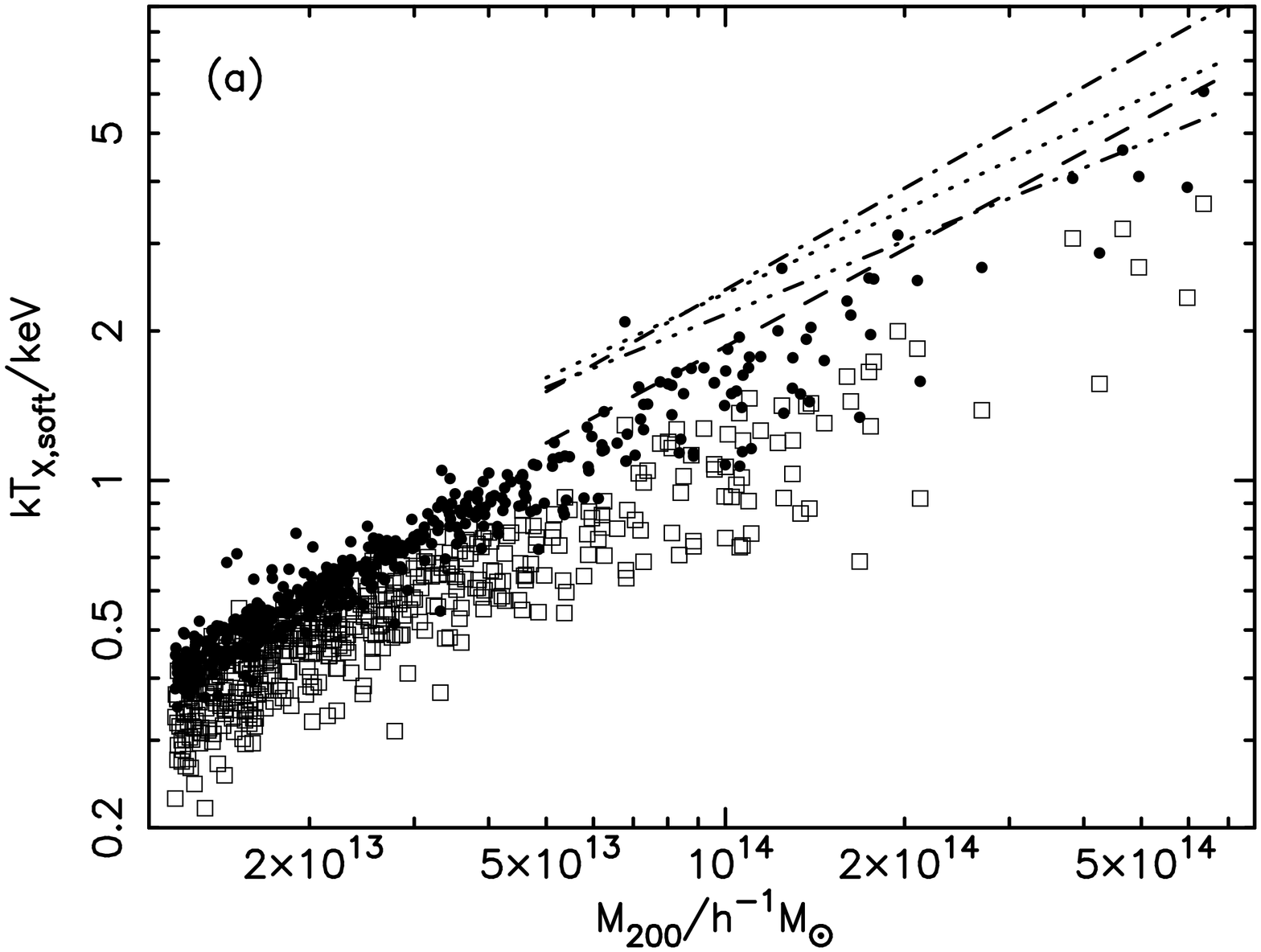,angle=270,width=8.5cm}
\psfig{file=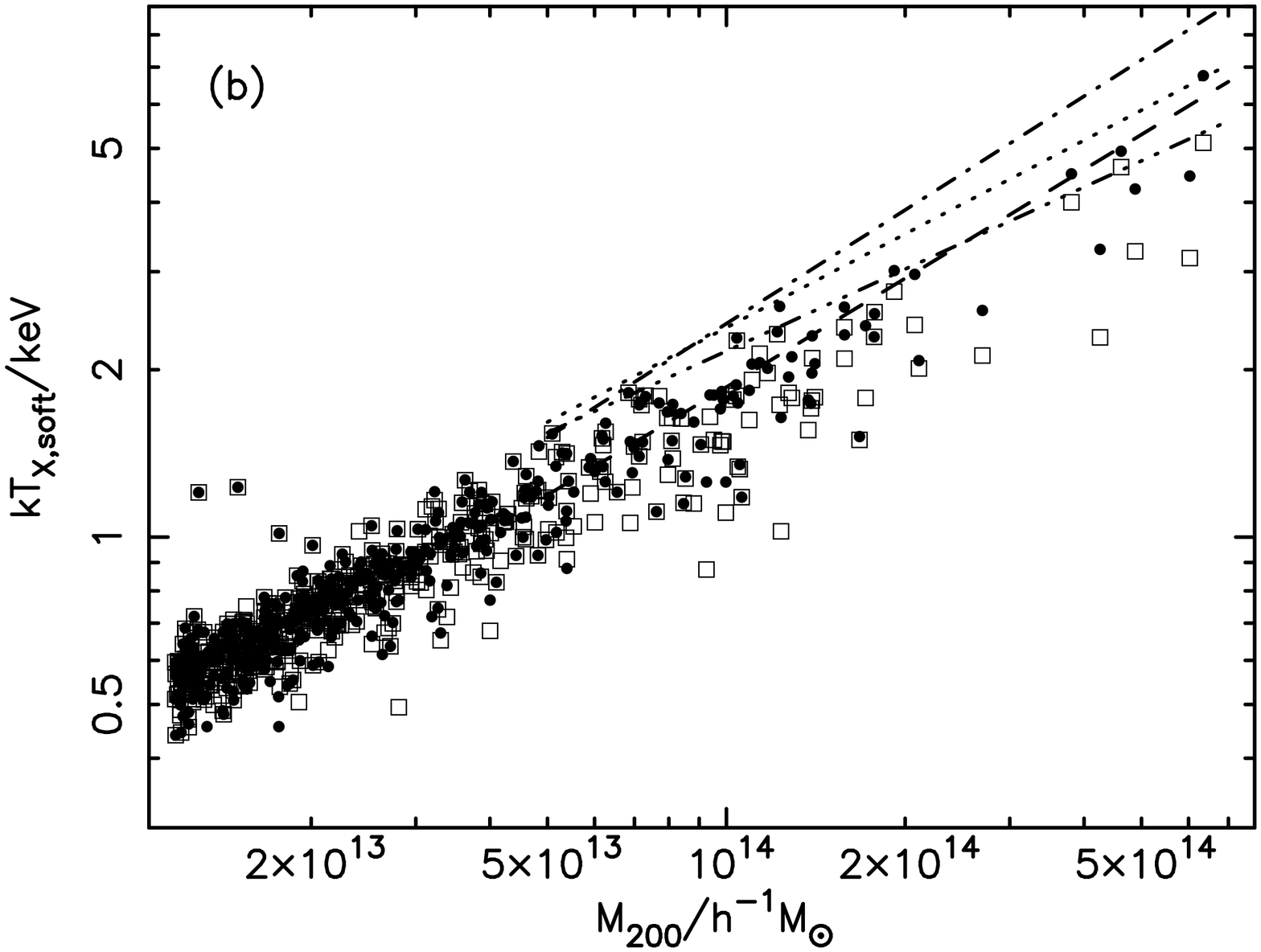,angle=270,width=8.5cm}
\psfig{file=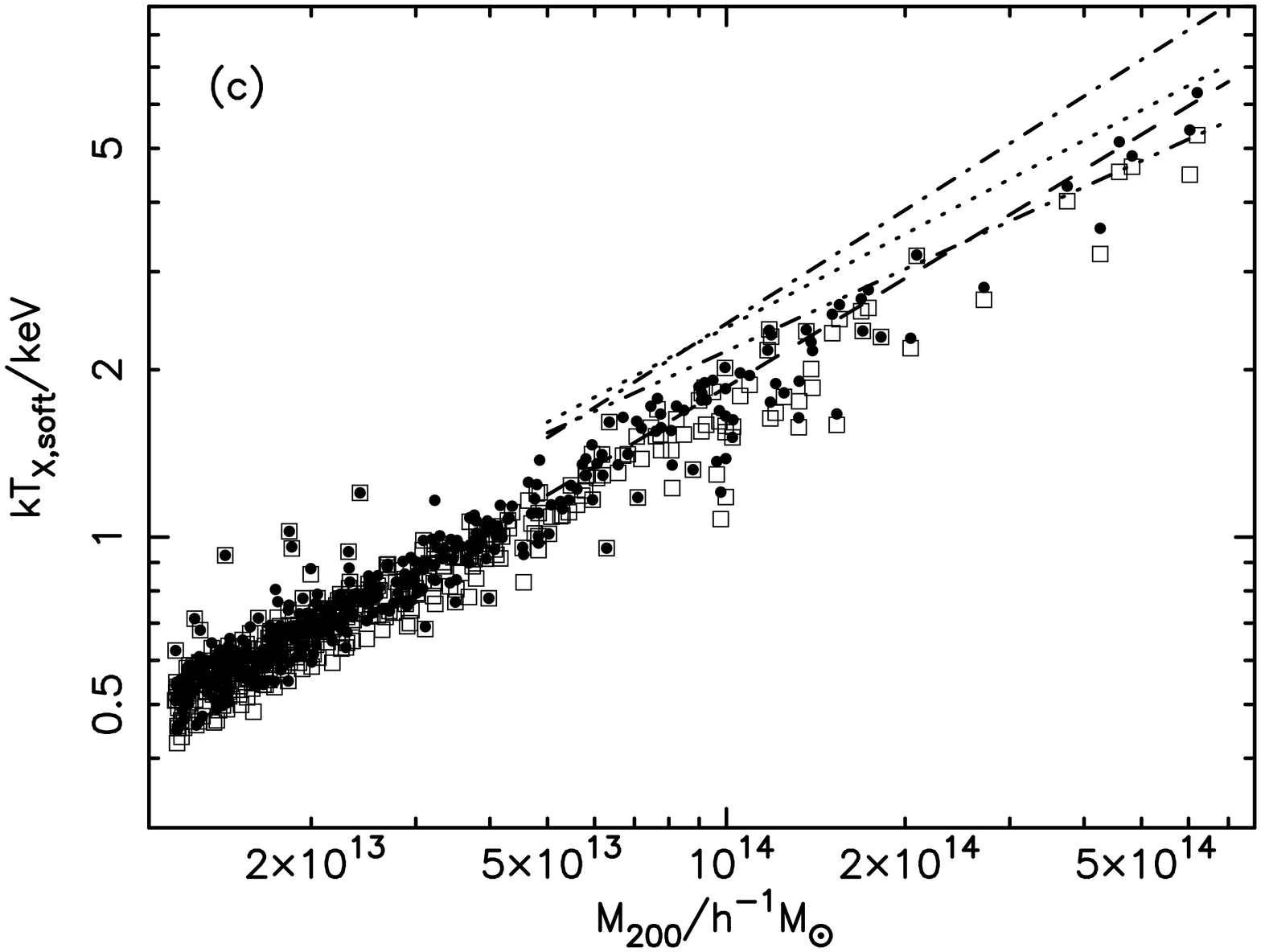,angle=270,width=8.5cm}
\caption{The X-ray temperature versus mass relation for (a) the {\it
Non-radiative}, (b) the {\it Radiative} and (c) the {\it Preheating}
simulations.  The open squares show the total soft-band X-ray
temperature whereas the filled circles exclude emission from (a) gas
with short cooling times, or (b,c) gas within the cooling radius.  The
different lines are from observations (Horner et al. 1999) using mass
estimates from galaxy velocity dispersions (dashed line), X-ray
temperature profiles (dot-dashed line), the isothermal-$\beta$ model
(dotted line), and emissivity profiles derived by surface brightness
deprojection (triple-dot-dashed line).}
\label{fig:tm}
\end{figure}

$kT_{X}-M_{200}$ relations for the clusters are presented in
Fig~\ref{fig:tm}.  We use the cooling table of Raymond \& Smith (1977)
to calculate the X-ray temperature in the soft band (0.3-1.5 keV), as
described in Section~\ref{eq:entropy}.  The open squares show the
temperature calculated using all the particles in the clusters.
However, the presence of cold, high-density gas in the cluster cores
(and also in infalling subclumps) gives emission-weighted
temperatures that are well below the virial temperature of the
cluster.  Accordingly, we also show as filled circles, for the {\it
Radiative} and {\it Preheating} runs, the `cooling-flow corrected
temperature' obtained by omitting emission from within the cooling
radius, as defined in Section~\ref{sec:coolflow}.  The change is most
important for high-mass clusters, for which it has the effect of both
tightening the relation and bringing it closer to the predicted slope
of two-thirds.  The cooling-flow correction does not work very well
for clusters in the {\it Non-radiative} simulation which have cool,
dense gas at all radii: for this run, therefore, we show instead the
effect of omitting all gas particles with cooling times shorter than
6\,Gyr.

Most of our clusters are smaller than those for which X-ray masses
have been determined.  Hence, to facilitate comparison with
observations, we fit a power-law to the temperature-mass relation for
clusters more massive than $5\times10^{13}\hMsol$.
Table~\ref{tab:tm} lists the normalisation, $A$, and slope,
$\alpha$ of the relation
\begin{equation}
kT_{X} = A \, (M_{200}/3\times 10^{14} h^{-1}{\rm M_{\odot}})^{1/\alpha} 
\, {\rm keV},
\label{eqn:tmrel}
\end{equation} 
where we have chosen to normalise at a mass scale
of $3\times10^{14}\hMsol$---towards the upper end of our
simulated catalogue but the lower-end of most observed ones.

We have divided the Table into three parts.  In the first, we list
results from previous simulations: EMN96---Evrard, Metzler \& Navarro
(1996); BN98---Bryan \& Norman (1998); T2001---Thomas et~al.~(2001);
ME01---Mathiesen \& Evrard (2001).  These use various cosmologies, but
fortunately the results do not seem to be very sensitive to this.
Much more important is the resolution of the simulation.  Thus ME01
have higher resolution than previous simulations (unfortunately they
do not state their precise mass-resolution in the paper) and find a
lower normalisation; our current simulations have a higher resolution
again and lower the normalisation still further.  The reason why the
increasing resolution lowers the emission-weighted temperature of the
gas is the presence of cold, dense gas in subclumps.  When we exclude
gas with a cooling time of less than 6\,Gyr from the calculation then
we find that the emission-weighted temperature rises once more to a
similar value to that found in the earlier, low-resolution simulations.
  
The middle section of the Table shows results for the clusters
described in the current paper for emission in the soft-band,
0.3--1.5\,keV, both with and without the inclusion of gas within the
cooling radius.  The cooling flow correction has little effect on
clusters in the {\it Non-radiative} simulation, mainly because of the
presence of cool gas in infalling subclumps (the effect of removing
all gas with short cooling times is much larger and was shown in the
first part of the Table).  On the other hand, gas with a short cooling
time in the {\it Radiative} and {\it Preheating} simulations resides
primarily in the cores of large clusters and its omission does
significantly raise the emission-weighted temperature.  In a previous
paper, Pearce et~al.~(2000), it was shown that radiative cooling
\emph{raises} the temperature of the of the intracluster medium and we
confirm that result.  Unfortunately, this is cancelled by the lower
temperature obtained by moving to higher resolution, so that the net
effect is to give temperature normalisations that are little changed
over earlier, non-radiative, low-resolution simulations

In the lower portion of Table~\ref{tab:tm}, we present some
observational determinations of cluster temperatures from:
HMF99---Horner, Mushotzky \& Scharf (1999); FRB01---Finoguenov,
Reiprich \& B\"ohringer (2001); XJW01---Xu, Jin \& Wu (2001).  It is
noticeable that various methods provide very different scaling
relations, both in normalisation and slope.  The two methods that
provide the best agreement with the simulations are those that combine
optical velocity dispersions either with {\it ASCA} temperatures or with
surface-brightness deprojection of {\it Einstein} data to create
emissivity profiles (White, Jones \& Forman 1997).  Unfortunately,
these are the least reliable as mass estimates from velocity
dispersions are prone to projection effects (e.g.~van Haarlem, Frenk
\& White 1997).  Also, the deprojection method requires a
large extrapolation from $\Delta\approx2000$ out to $\Delta=200$.
The highest normalisation is provided by using resolved temperature
profiles.  In principle this should be the most accurate method but as
yet the temperature profiles are poorly-determined and a high degree of
modelling is required.  In two papers by Nevalainen, Markevitch \&
Forman (1999, 2000), for example, the enclosed gas mass fraction in
the clusters A\,401 and A\,3571 can be seen to be steeply rising at
the virial radius, contrary to expectation.  

The greatest degree of consensus is given by different authors using
the isothermal-$\beta$ model, which is another way of saying that
different groups measure the same relationship between $\beta_{\rm
fit}$ and temperature.  In this model, the mass profile is
\begin{equation}
M(<r) \approx 1.11\times10^{14}\beta_{\rm fit}\,{kT\over\keV}\,
         {r^2\over r_c^2+r^2}\,
         \left(r\over h^{-1}\Mpc\right)\,\hMsol,
\end{equation}
which gives a mass within $r_{200}$ of
\begin{equation}
M_{200} \approx 7.69\times10^{13}
        \left(\beta_{\rm fit}\,kT\over\keV\right)^{3\over2}
        \left(r_{200}^2\over r_c^2+r_{200}^2\right)^{3\over2}\hMsol.
\label{eq:tmisobeta}
\end{equation}
In most cases $r_{200}\gg r_c$ so that the correction term in the above
equation for the finite core radius is approximately unity.

\begin{table}
\caption{Power-law fits to the observational and simulated
temperature-mass relations of X-ray clusters: cluster sample; slope of
relation, $\alpha$; value of $kT/$keV at $3\times10^{14}\hMsol$, $A$.}
\label{tab:tm}
\begin{tabular}{llcc}
\hline
Sample&& $\alpha$& $A$ \\
\hline
EMN96              & soft band             & 1.50& 3.6\\
BN98               & bolometric            & 1.50& 2.9\\
T2001              & bolometric            & 1.50& 3.6\\
ME01               & broad band, high res. & 1.51& 2.6\\
{\it Non-radiative}& bolometric,                        & 1.50& 2.1\\
                   & bolometric, $t_{\rm cool}>6$\,Gyr  & 1.51& 3.3\\
\hline
{\it Non-radiative}& soft-band             & 1.69& 2.0\\
                   & cooling-flow corrected& 1.69& 2.2\\
{\it Radiative}    & soft-band             & 1.96& 2.8\\
                   & cooling-flow corrected& 1.64& 3.4\\
{\it Preheating}   & soft-band             & 1.67& 3.2\\
                   & cooling-flow corrected& 1.61& 3.5\\
\hline
HMS99              & velocity dispersions  & 1.53& 3.8\\
                   & temperature profiles  & 1.48& 5.1\\
                   & emissivity profiles   & 2.06& 3.7\\
                   & isothermal-$\beta$ model & 1.78& 4.4\\
FRB01              & isothermal-$\beta$ model & 1.67& 4.1\\
                   & polytropic-$\beta$ model, low-$T$ & 1.87& 4.3\\
                   & polytropic-$\beta$ model, high-$T$ & 1.48& 4.4\\
XJW01              & NFW-model             & 1.81& 4.9\\
                   & isothermal-$\beta$ model & 1.60& 4.3\\
\hline
\end{tabular}
\end{table}

The most extensive analysis of this kind was performed by Finoguenov,
Reiprich \& B\"ohringer (2001) for the HIFLUGCS (Highest X-ray Flux
Galaxy Cluster Sample) from the ROSAT All-Sky Survey.  They used
$\beta_{\rm fit}$ values taken from fits to the ROSAT PSPC data and
temperatures mostly from ASCA. Where information on temperature
gradients was available they generalised the $\beta$-model to allow a
polytropic Equation of State and this gave very similar results.  A
similar result is found by Xu, Jin \& Wu (2001) using a smaller sample
and by Horner, Mushotzky \& Scharf (1999) using ASCA data on 38
clusters from Fukazawa (1997).  The slopes of the observed
temperature-mass relations are slightly steeper than 1.5,
reflecting the fact that the measured values of $\beta_{\rm fit}$ are
a slowly increasing function of mass, as we found for the simulated
clusters in Section~\ref{sec:sbprof}.

\begin{figure}
\psfig{file=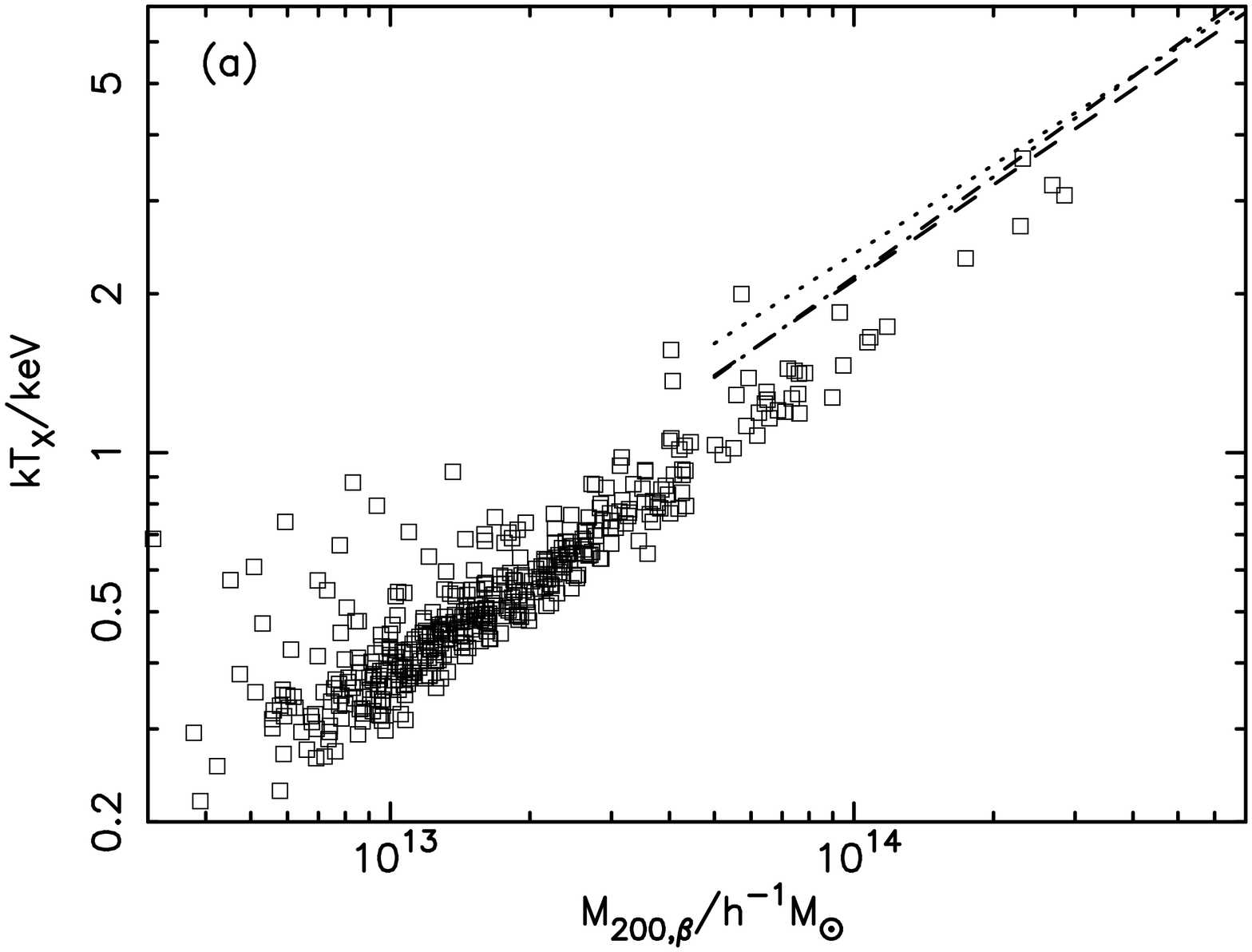,angle=270,width=8.7cm}
\psfig{file=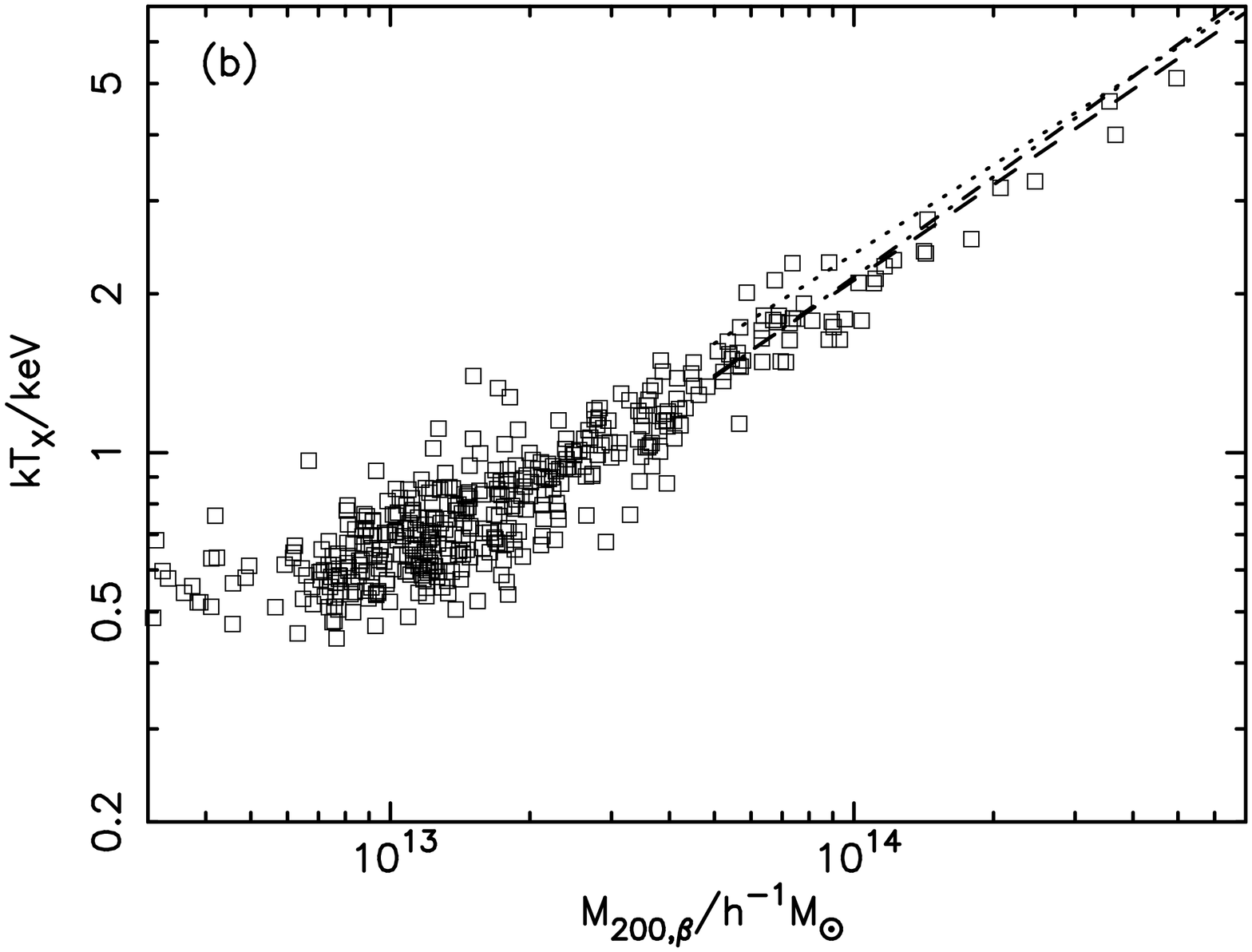,angle=270,width=8.7cm}
\psfig{file=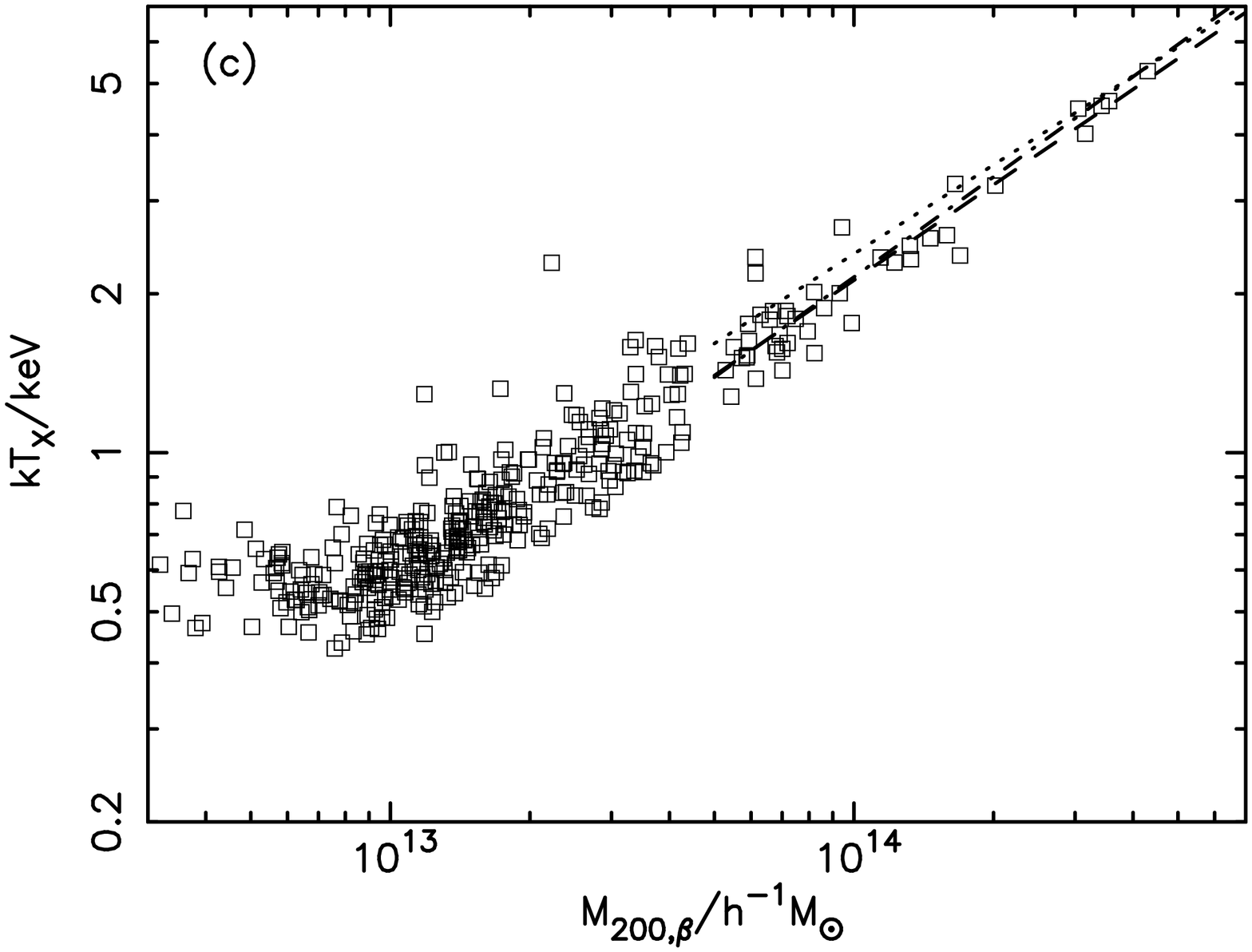,angle=270,width=8.7cm}
\caption{The X-ray temperature versus mass relation for (a) the {\it
Non-radiative}, (b) the {\it Radiative} and (c) the {\it Preheating}
simulations, where the mass is estimated from the isothermal-$\beta$ model.  
The dotted, dashed and dot-dashed lines are observed masses using
isothermal-$\beta$ model from HMS99, FRB01 and XJW01, respectively.}
\label{fig:tmbeta}
\end{figure}

To test whether the temperature-mass relation using the
isothermal-$\beta$ model is consistent with the observations, we
estimate the masses of the simulated clusters using
equation~\ref{eq:tmisobeta} and the values of $\beta_{\rm fit}$ from
Section~\ref{sec:sbprof}.  The resulting scaling relation, shown in
Figure~\ref{fig:tmbeta}, is fully consistent with the observations.
We do not want to over-interpret this result as the method of
determining $\beta_{\rm fit}$ in Section~\ref{sec:sbprof} is
far removed from the analysis that is carried out on real X-ray data.
Ideally, one would create mock observations from the simulations and
analyse them in the same way but that is a complex procedure that is
beyond the scope of the present paper.  Nevertheless, we tentatively
conclude that the isothermal-$\beta$ model underestimates cluster
masses and that there is no disagreement between the masses of
simulated and real clusters.

\subsection{Luminosity--Temperature relation}

We define the bolometric X-ray luminosity, estimated from emission in the
soft band as 
\begin{equation}
L_X={\Lambda_{\rm bol}(T_X)\over\Lambda_{\rm soft}(T_X)}
\sum_i{m_i\rho_i\Lambda_{\rm soft}(T_i,Z)\over(\mu m_{\rm H})^2},
\label{eq:lx}
\end{equation}
where $\mu m_{\rm H}=10^{-24}$g is the mean molecular mass, $T_X$ is
the soft-band X-ray temperature as defined in equation~(\ref{eq:ktx}),
and $\Lambda_{\rm bol}$ and $\Lambda_{\rm soft}$ are the bolometric
and soft-band cooling functions, respectively.  

\begin{figure}
\psfig{file=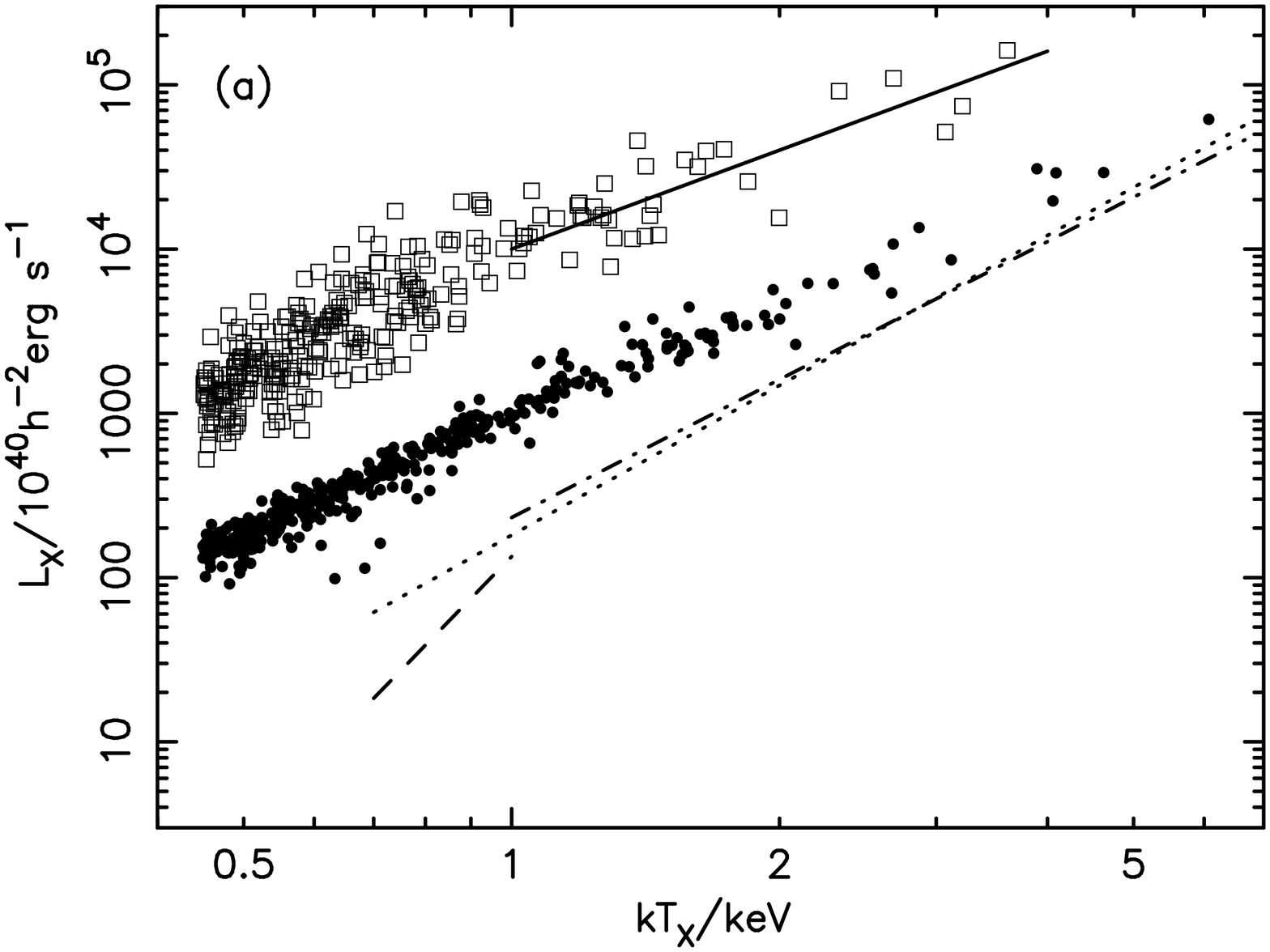,angle=270,width=8.7cm}
\psfig{file=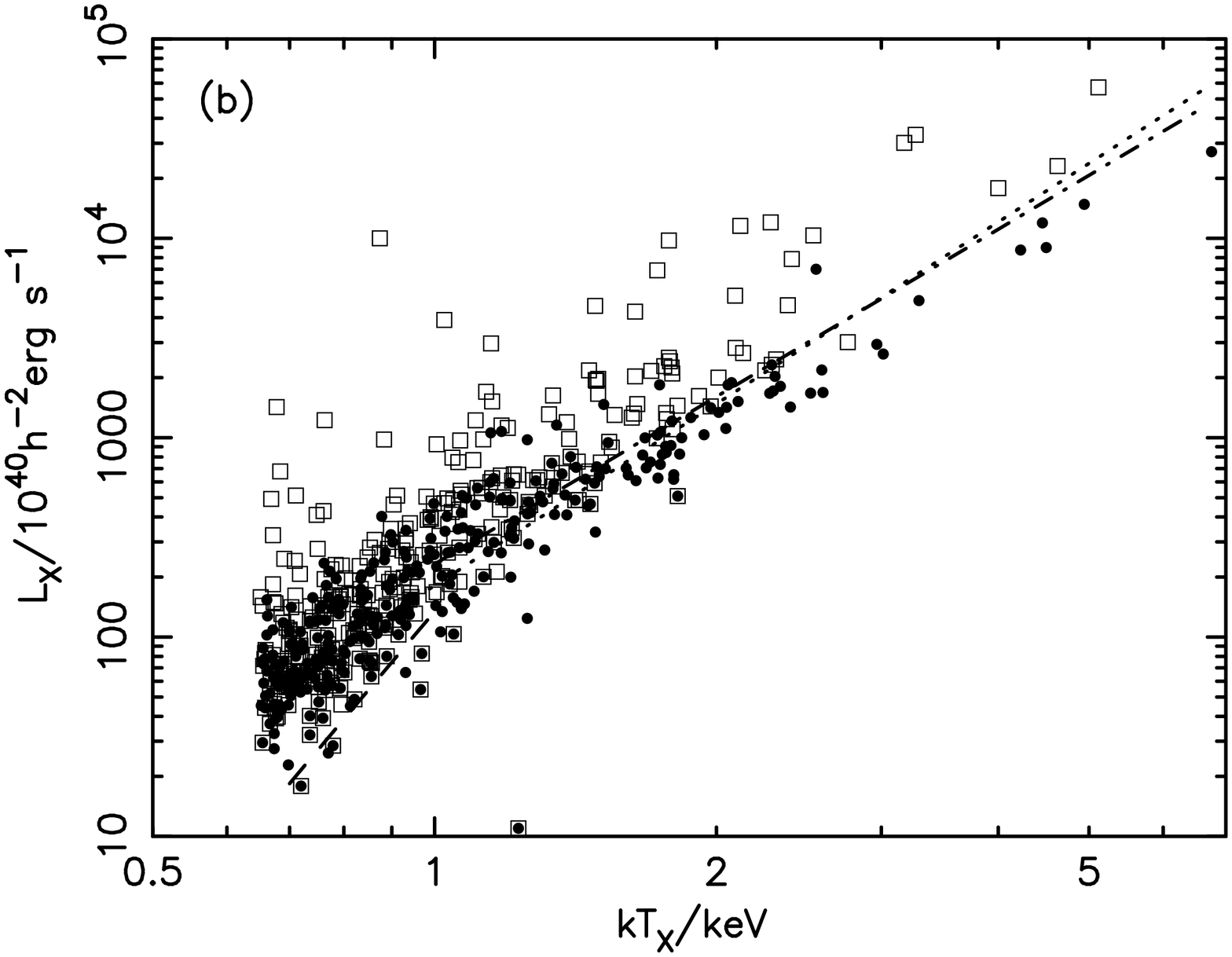,angle=270,width=8.7cm}
\psfig{file=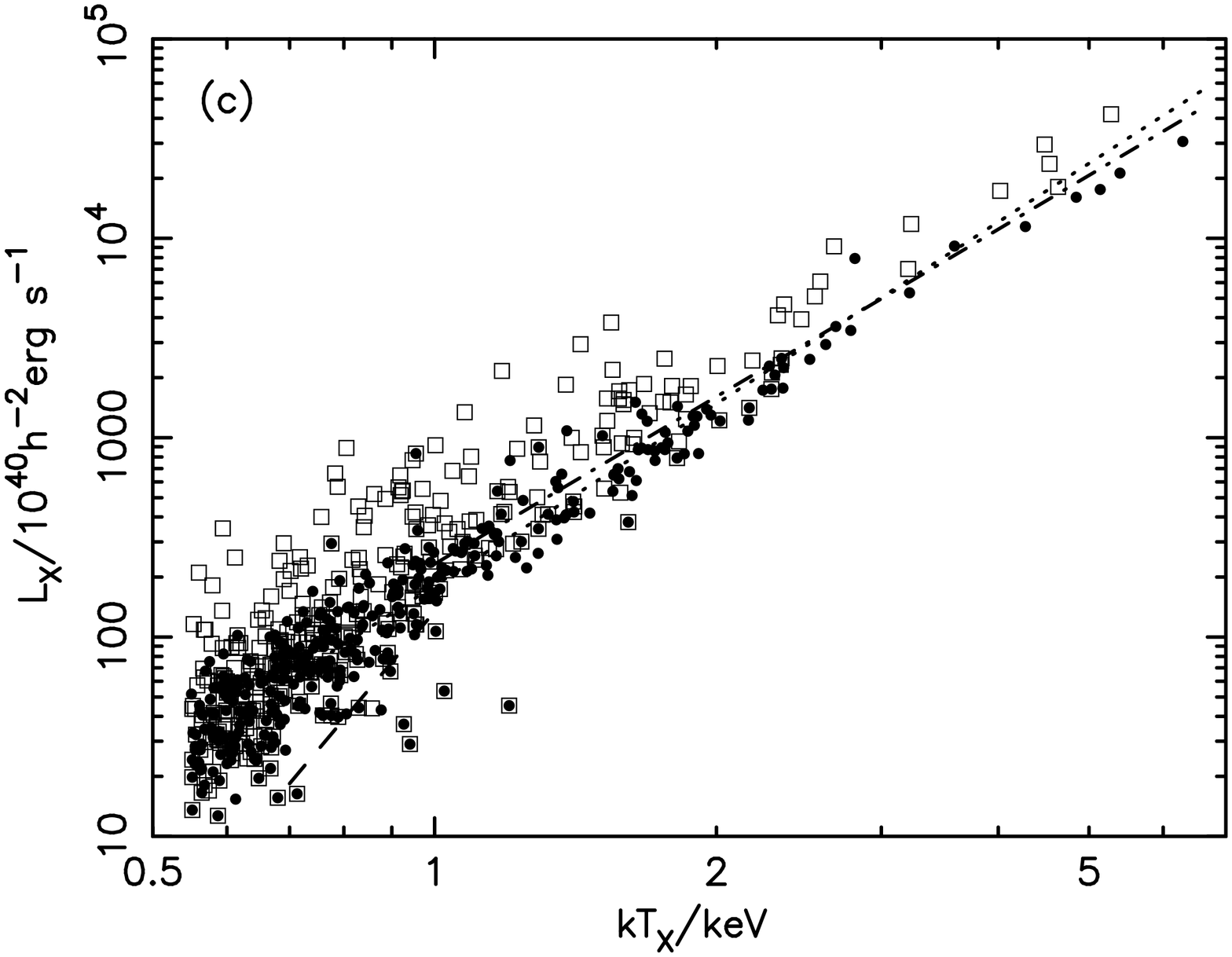,angle=270,width=8.7cm}
\caption{The bolometric X-ray luminosity vs. temperature as estimated
from the soft band for (a) the {\it Non-radiative}, (b) the {\it
Radiative} and (c) the {\it Preheating} simulations.  The open squares
use the total soft-band emission, whereas the filled circles
exclude emission from (a) gas with short cooling times, or (b,c) gas
within the cooling radius.  The dashed, dot-dashed and dotted lines
are observed relations from Xue \& Wu (2000); the solid line in panel
(a) shows the self-similar relation $L_X\propto T_X^2$.}
\label{fig:lxtx}
\end{figure}

$L_X$ is plotted against $T_X$ both with and without emission from
within the cooling radius in Figure~\ref{fig:lxtx}.  We have
restricted the temperature ranges to those for which the catalogues
are reasonably complete (as judged by looking at the upper locus of
the points in Figure~\ref{fig:tm}).
Also shown on the Figure are observed relations from Xue \& Wu (2000)
who provided a compilation of observed X-ray temperatures and
luminosities from the literature. Their sample was divided into 3
subsamples: groups (below 1 keV), clusters (above 1 keV) and the
mixture of the two. Their best fit of each category is shown as the
dashed, dot-dashed and dotted lines in the Figure.

The first thing to note is that the luminosities of the clusters in
the {\it Non-radiative} simulation are much greater than for observed
clusters.  These are much reduced, however, by the removal of gas with
short cooling times.  At temperatures above 1\,keV, the uncorrected
relation follows the self-similar relation $L_X\propto T_X^2$ expected
for bremsstrahlung radiation.  At lower temperatures, the luminosity
might be expected to exceed the self-similar prediction because of the
added flux from line emission, but in fact it is reduced.  This is
because the cores of the smaller clusters are relatively less well
resolved.  We do not regard this lack of resolution as important
because the gas in the core has a short cooling time and contributes a
negligible amount to the total emission in the {\it Radiative} and
{\it Preheating} simulations.

Both the {\it Radiative} and {\it Preheating} simulations show
$L_X$-$T_X$ relations that lie much closer to the observations.  The
agreement is best for $kT_X>$1\,keV, less so at lower temperatures.
This may be because we have not raised the entropy sufficiently in the
cores of these systems.  However, the observational determination of
the X-ray luminosity of low-temperature clusters is very hard (see
e.g.~Helsdon \& Ponman 2000; Wu \& Xue 2002a; Voit et al.~2002) and so
the lack of agreement is not so serious.  For both simulations, but
most especially for the {\it Radiative} one, omission of gas within
the cooling radius vastly reduces the scatter and brings the outliers
down to the main relation.

\section{Conclusions}
We have analysed the properties of clusters drawn from three
$N$-body, hydrodynamical simulations of the $\Lambda$CDM cosmology.
Each uses the same initial conditions but varies in its treatment of
the gas physics: a standard adiabatic ({\it Non-radiative}) model, a {\it
Radiative} that includes radiative cooling of the gas, and a {\it
Preheating} model that also includes cooling but in addition
impulsively heats the gas prior to cluster formation.  Each
simulation generated over 500 clusters, complete in mass down to 
$1.18\times10^{13} h^{-1}\Msol$.

The {\it Non-radiative} simulation does not reproduce the observations
but was used as a test of the simulation procedure.  The clusters
drawn from this simulation show no signs of numerical heating and
behave self-similarly in their properties.

Both the {\it Radiative} and the {\it Preheating} simulations
reproduce three key observational relations:
\begin{itemize}
\item The entropy in the cores of low-mass clusters lies above the
self-similar relation.  The measured value at 0.1\,$r_{\rm vir}$\ tends
towards a value of approximately 100\,$h^{-1/3}$keV\,cm$^2$ at low
masses, with very large scatter.
\item The luminosity-temperature relations are much reduced in
normalisation relative to the {\it Non-radiative} simulation, and lie
close to the observed relation above 1\,keV once corrected for
cooling-flow emission.  At lower temperatures we still seem to
overpredict the X-ray luminosity, although the observational errors are
large.
\item We have shown in an earlier paper (Thomas et~al.~2002) that the
temperature-mass relation in the inner parts of clusters, within
$r_{2500}$, agrees with observations.  In this paper we reproduce
earlier results that show that simulated cluster masses within
$r_{200}$ are significantly greater than observed ones for a given
cluster temperature.  However, we show that the use of the
isothermal-$\beta$ and related models can lead to an underestimate of
cluster masses and once this is taken into account the observations
and simulations are once again brought into agreement.  The
implications of this for the determination of $\sigma_8$ from the
observed cluster temperature function are the subject of a separate
paper, in preparation.
\end{itemize}

The basic explanation for the agreement in the properties of simulated
and observed clusters is an increase in entropy in the cluster cores,
over and above that expected in an adiabatic simulation.  In the {\it
Radiative} simulation this occurs via the removal of low-entropy gas
by radiative cooling, whereas in the {\it Preheating} simulation it
comes about through the imposed energy increase at high redshift.
These two mechanisms differ considerably in the amount of cooled gas
that result: at the end of the simulations the global cooled baryon
fractions are 15 per cent and 0.4 per cent respectively, bracketing
the observed value.  Thus, while neither model is a correct
description of clusters, one might expect that the true model gives
rise to similar entropy profiles.

We showed that the mass-deposition rate in cooling flows, i.e.~the
amount of gas cooling to low temperatures in the cluster cores, is
reasonably well-approximated by the usual method of dividing the
luminosity by the enthalpy within the cooling radius within which the
mean cooling time is equal to 6\,Gyr.  However, the actual
mass-deposition rate is highly stochastic and may be driven by the
infall of high-density subclumps.  Higher resolution simulations are
required to investigate this further.

In the simulations that we have described in this paper, the cooling
is limited by the numerical resolution.  Future simualtions will move
to higher resolution and will have to include the feedback of energy
from supernovae.  This will act as a form of preheating, thus removing
the ad-hoc nature of the current model, although it may be that other
heating mechanisms are also required.  We fully expect that
realistic models will emerge that successfully replicate all the
observed features of the intracluster medium.

\section*{Acknowledgments}

The simulations described in this paper were carried out on the
Cray-T3E at the Edinburgh Parallel Computing Centre as part of the
Virgo Consortium investigations of cosmological structure formation.
OM is supported by a DPST Scholarship from the Thai government; PAT is
a PPARC Lecturer Fellow.

\bsp

\label{lastpage}

\end{document}